%% file: main.tex
\documentclass[sigconf]{acmart}

\usepackage{array}
\usepackage{multirow}
\usepackage{glossaries}
\usepackage{enumitem}
\usepackage{xcolor}
\usepackage{placeins}

\newcommand{\hl}[1]{{#1}}
\newcommand{\hhl}[1]{{#1}}

\AtBeginDocument{%
  }

\copyrightyear{2025}
\acmYear{2025}
\setcopyright{acmlicensed}
\acmConference[CHI '25]{CHI Conference on Human Factors in Computing Systems}{April 26-May 1, 2025}{Yokohama, Japan}
\acmBooktitle{CHI Conference on Human Factors in Computing Systems (CHI '25), April 26-May 1, 2025, Yokohama, Japan}
\acmDOI{10.1145/3706598.3713496}
\acmISBN{979-8-4007-1394-1/25/04}

\newacronym{sys}{SYS}{DanmuA11y}

\begin{document}

\title[DanmuA11y: Making Danmu Accessible to Blind and Low Vision Viewers]{DanmuA11y: Making Time-Synced On-Screen Video Comments (Danmu) Accessible to Blind and Low Vision Users via Multi-Viewer Audio Discussions}

\author{Shuchang Xu}
\affiliation{
\institution{Hong Kong University of Science and Technology}
\city{Hong Kong}
\country{China}
}
\orcid{0000-0002-7642-9044}
\email{sxuby@connect.ust.hk}

\author{Xiaofu Jin}
\affiliation{
\institution{Hong Kong University of Science and Technology}
\city{Hong Kong}
\country{China}
}
\orcid{0000-0002-7239-3769}
\email{xjinao@connect.ust.hk}

\author{Huamin Qu}
\affiliation{
\institution{Hong Kong University of Science and Technology}
\city{Hong Kong}
\country{China}
}
\orcid{0000-0002-3344-9694}
\email{huamin@cse.ust.hk}

\author{Yukang Yan}
\authornote{This is the corresponding author.}
\affiliation{
\institution{University of Rochester}
\city{New York}
\country{United States}
}
\orcid{0000-0001-7515-3755}
\email{yukang.yan@rochester.edu}

\begin{abstract}
  \input{Sections/0_Abstract}
\end{abstract}



\begin{teaserfigure}
  \includegraphics[width=\textwidth]{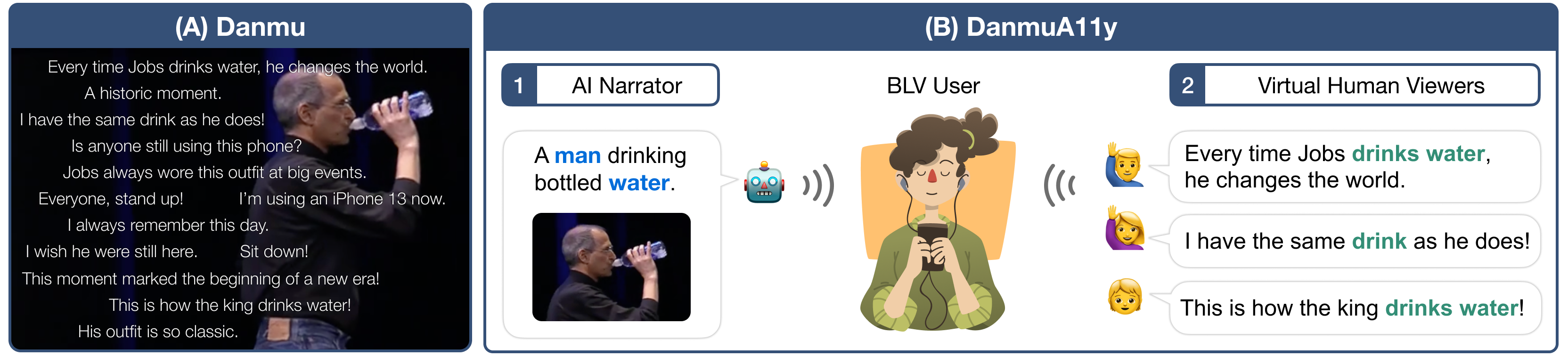}
  \caption{(A) Danmu is a video-commenting feature that overlays time-synced user comments onto videos. It creates a co-watching experience for online viewers. However, its visual-centric design poses significant challenges for blind and low vision (BLV) viewers. (B) DanmuA11y makes Danmu accessible to BLV viewers by transforming it into \textbf{multi-viewer audio discussions}. It includes two types of virtual viewers: (1) an AI narrator offering visual descriptions, and (2) virtual human viewers verbalizing curated Danmu comments. These audio discussions are seamlessly integrated into the video, creating an enjoyable and socially engaging experience for BLV viewers. (All Danmu comments in this paper were originally in Mandarin and have been translated into English.)}
  \Description{This figure contains two images. The left image shows Danmu, a video commenting feature that overlays a large number of time-synced text comments on the video screen. This feature creates a co-watching experience for online viewers but poses significant challenges for blind and low-vision viewers. The right image shows DanmuA11y, a system that makes Danmu accessible to BLV viewers by transforming Danmu into multi-viewer audio discussions. At the center of the image is a user with blindness. To the user's left is an AI narrator that reads out visual descriptions of the video keyframe: "A man drinking bottled water." To the user's right is a group of virtual human viewers who verbalize curated Danmu comments. The first viewer says: "Every time Jobs drinks water, he changes the world." The second viewer says: "I have the same drink as he does!" The third viewer says: "This is how the king drinks water!" These audio discussions are seamlessly integrated into the video timeline, creating an enjoyable and socially engaging experience for BLV viewers.}
  \label{fig:teaser}
\end{teaserfigure}

\maketitle

\input{Sections/1_Introduction}

\input{Sections/2_RelatedWork}

\input{Sections/3_FormativeStudy}

\input{Sections/4_System}
\input{Sections/6_UserEvaluation}
\input{Sections/7_Discussion}

\input{Sections/8_Conclusion}

\begin{acks}
The authors would like to thank the participants for their support during the studies. We also thank the reviewers for their constructive feedback. We are grateful to Ciyuan Yang, Liwenhan Xie, Ke Wang, and Yan Cui for their insightful discussions. Furthermore, we thank Chang Chen, Zichen Liu, Jindu Wang, Wenshuo Zhang, Runhua Zhang, and Fanseu Barakeel for their help in proofreading the manuscript. This work is partially supported by the Research Grants Council of the Hong Kong Special Administrative Region under Hong Kong PhD Fellowship Scheme with No. PF23-94000.
\end{acks}

\bibliographystyle{ACM-Reference-Format}
\bibliography{references}

\appendix
\newpage\onecolumn

\section{Supplementary Tables and Figures}
\begin{figure}[htbp]
    \centering
    \includegraphics[width=0.48\textwidth]{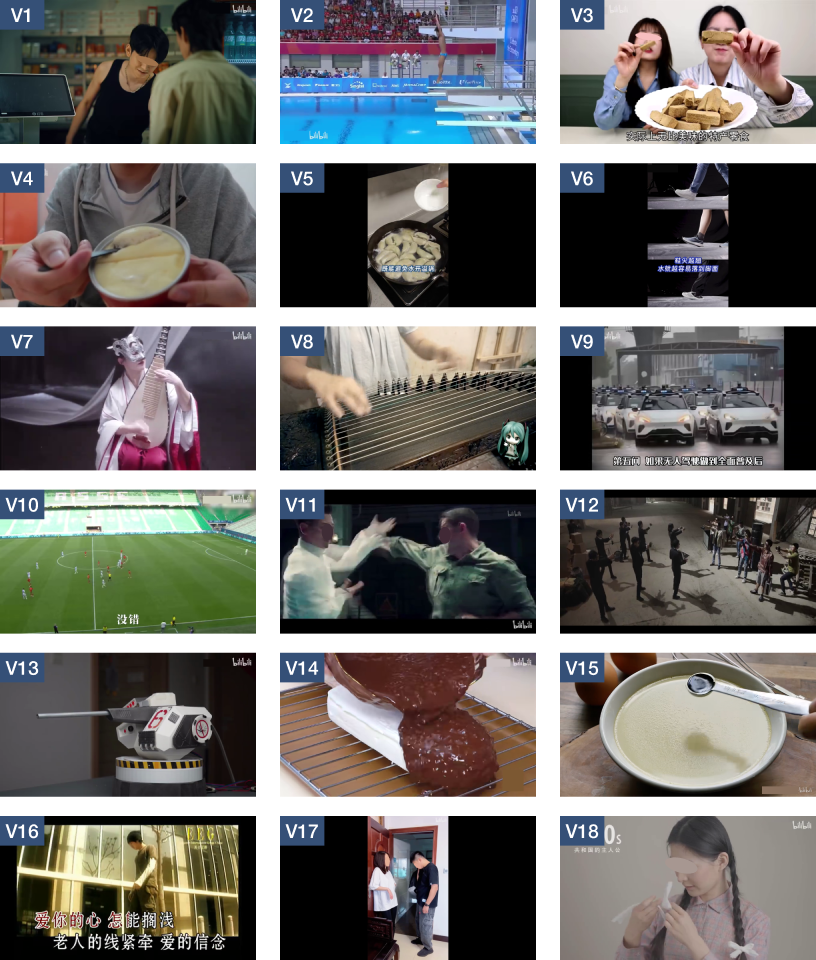}
    \caption{Screenshots from the 18-video dataset. Video creators' IDs and human faces are obscured for privacy.}
    \Description{This figure shows the video screenshots for the eighteen videos used in the evaluation study.}\label{fig:video_screenshots}
\end{figure}

This section provides the following tables and figures.
\begin{itemize}
    \item Table~\ref{tab:demographics} shows participants' demographics.
    \item Table~\ref{tab:video_formative} lists the six videos used in the formative study.
    \item Table~\ref{tab:video_evaluation} and Figure~\ref{fig:video_screenshots} provide descriptions and screenshots of the 18 videos used in the evaluation study.
    \item Table~\ref{tab:subjective_ratings} presents detailed statistics on the subjective ratings from the evaluation study.
    \item Table~\ref{tab:trials} outlines the uncompleted trials in the controlled comparison.
\end{itemize}

\input{Tables/demographics}
\input{Tables/video_formative}

\input{Tables/video_evaluation}

\input{Tables/evaluation_stats}
\input{Tables/trials}

\end{document}

%% file: Sections/0_Abstract.tex
By overlaying time-synced user comments on videos, Danmu creates a co-watching experience for online viewers. However, its visual-centric design poses significant challenges for blind and low vision (BLV) viewers. Our formative study identified three primary challenges that hinder BLV viewers' engagement with Danmu: the lack of visual context, the speech interference between comments and videos, and the disorganization of comments. To address these challenges, we present DanmuA11y, a system that makes Danmu accessible by transforming it into multi-viewer audio discussions. DanmuA11y incorporates three core features: (1) \textit{Augmenting Danmu with visual context}, (2) \textit{Seamlessly integrating Danmu into videos}, and (3) \textit{Presenting Danmu via multi-viewer discussions}. Evaluation with twelve BLV viewers demonstrated that DanmuA11y significantly improved Danmu comprehension, provided smooth viewing experiences, and fostered social connections among viewers. We further highlight implications for enhancing commentary accessibility in video-based social media and live-streaming platforms.

%% file: Sections/1_Introduction.tex
\section{Introduction}

Online video platforms have become important channels for viewers to exchange thoughts, emotions, and knowledge related to videos \cite{liu2021makes,wu2018danmaku}. Traditionally, these interactions occur in a separate comments section, detached from the video itself, as exemplified by platforms like YouTube\footnote{https://www.youtube.com/}. This structure often fails to capture viewers' immediate reactions to specific video moments, leading to a disconnect between the video and the commentary \cite{ma2017video}. To address this limitation, several Asian video platforms have introduced a video-commenting feature known as Danmu. It enables viewers to post comments that are synchronized with the video timeline. These \textit{time-synced} comments are \textit{visually overlaid} on the video screen (see Figure~\ref{fig:teaser} (A)), allowing viewers to see others' instant reactions as they watch the video. Moreover, viewers can interact with each other by responding to previous comments \cite{ma2017video}. Consequently, Danmu creates a socially engaging experience that closely simulates the sensation of co-watching videos with other viewers \cite{ma2017video,chen2017watching,sun2016videoforest}.

However, Danmu is primarily designed for \textit{visual consumption}. Danmu comments are visually overlaid onto videos, often in overwhelming volumes \cite{chen2022danmuvis,ma2017video,cao2023visdmk}. These comments are scattered across the screen, requiring viewers to visually process multiple simultaneous comments and discern social interactions, such as identifying which comments are replies to others. These visual-centric characteristics pose significant challenges for blind and low vision (BLV) viewers, making it difficult for them to interpret Danmu comments and socially connect with other viewers.

To understand the practices and challenges of BLV viewers in accessing Danmu, we conducted formative interviews and co-watching sessions with eight BLV participants who regularly watched videos on Danmu-enabled platforms. Participants accessed Danmu using a design probe modeled after the most effective tool they currently use: an auto-scrolling list of Danmu comments. 
Our research revealed three significant challenges that hinder their engagement with Danmu: 
First, Danmu comments often discuss visual elements that are not accessible to BLV viewers. This \textbf{lack of visual context} makes it difficult for them to understand the discussion topics. 
Second, while BLV viewers use screen readers to access Danmu, the audio from screen readers often overlaps with the video's speech. This \textbf{speech interference} prevents them from enjoying both content at the same time. 
Third, while Danmu utilizes screen space to display a large number of comments simultaneously, they become disorganized when accessed sequentially via screen readers. This \textbf{disorganization of comments} makes it tedious for BLV viewers to follow audience discussions and socially engage with other viewers.

To address these challenges, we present DanmuA11y, a system that makes Danmu accessible by transforming Danmu into multi-viewer audio discussions. 
DanmuA11y encompasses three core features: (1) \textit{\textbf{Augmenting Danmu with Visual Context}}: DanmuA11y supplements Danmu comments with descriptions of the visual context, allowing BLV viewers to easily grasp the discussion topics; 
(2) \textit{\textbf{Seamlessly Integrating Danmu into Videos}}: By optimizing the insertion timing of Danmu comments in the video, DanmuA11y enables viewers to enjoy both content without speech overlap; and 
(3) \textit{\textbf{Presenting Danmu via Multi-Viewer Discussions}}: To create a co-watching experience, DanmuA11y organizes Danmu comments into dialogues and 
uses spatial audio to simulate the sensation of other viewers conversing around the user. 
Powered by these features, DanmuA11y offers an enjoyable and socially engaging experience for BLV video viewers.

To evaluate DanmuA11y, we conducted a within-subject study with 12 BLV participants, who compared DanmuA11y to a baseline system that simulated their current practices. Each participant watched two similar groups of videos using two systems, respectively. The results showed that DanmuA11y significantly improved Danmu comprehension ($p<.01$), provided a smoother viewing experience ($p<.01$), and enhanced the sense of co-watching with others ($p<.01$) compared to the baseline. Participants reported that DanmuA11y made video watching more enjoyable and socially engaging, with a strong willingness to use it in the future ($\mu=6.92$ on a seven-point Likert scale). 
Based on our findings, we discussed directions for personalizing DanmuA11y, and summarized implications for improving commentary accessibility in broader contexts, such as video-based social media and live-streaming platforms.

Our contributions are threefold:
\begin{itemize}
    \item We identify the motivations, needs, current practices, and challenges of BLV viewers in accessing Danmu, derived from a formative study;
    \item We present DanmuA11y, a system that makes Danmu accessible to BLV viewers through three core features: (1) augmenting Danmu with visual context, (2) seamlessly integrating Danmu into videos, and (3) presenting Danmu via multi-viewer audio discussions;
    \item We contribute an evaluation study that demonstrates how BLV viewers engage with Danmu using DanmuA11y, and derive design implications to improve commentary accessibility in video and live-streaming platforms.
\end{itemize}

%% file: Sections/2_RelatedWork.tex
\section{Background and Related Work}
Our work extends prior research in three areas: (1) Danmu and its accessibility, (2) social media accessibility, and (3) video accessibility for BLV viewers.

\subsection{Danmu and its Accessibility}
Danmu, also known as Danmaku, is a video-commenting feature that displays time-synced user comments as moving subtitles overlaid on the video screen. Originating in Japan, this feature has gained widespread popularity on Asian video platforms. For example, in China, nearly all major online video platforms support Danmu \cite{chen2017watching}. As of 2024, China's first Danmu-enabled video platform, Bilibili\footnote{https://www.bilibili.com/}, has accumulated over 20 billion Danmu comments.

\begin{figure}[ht]
    \centering
    \includegraphics[width=0.47\textwidth]{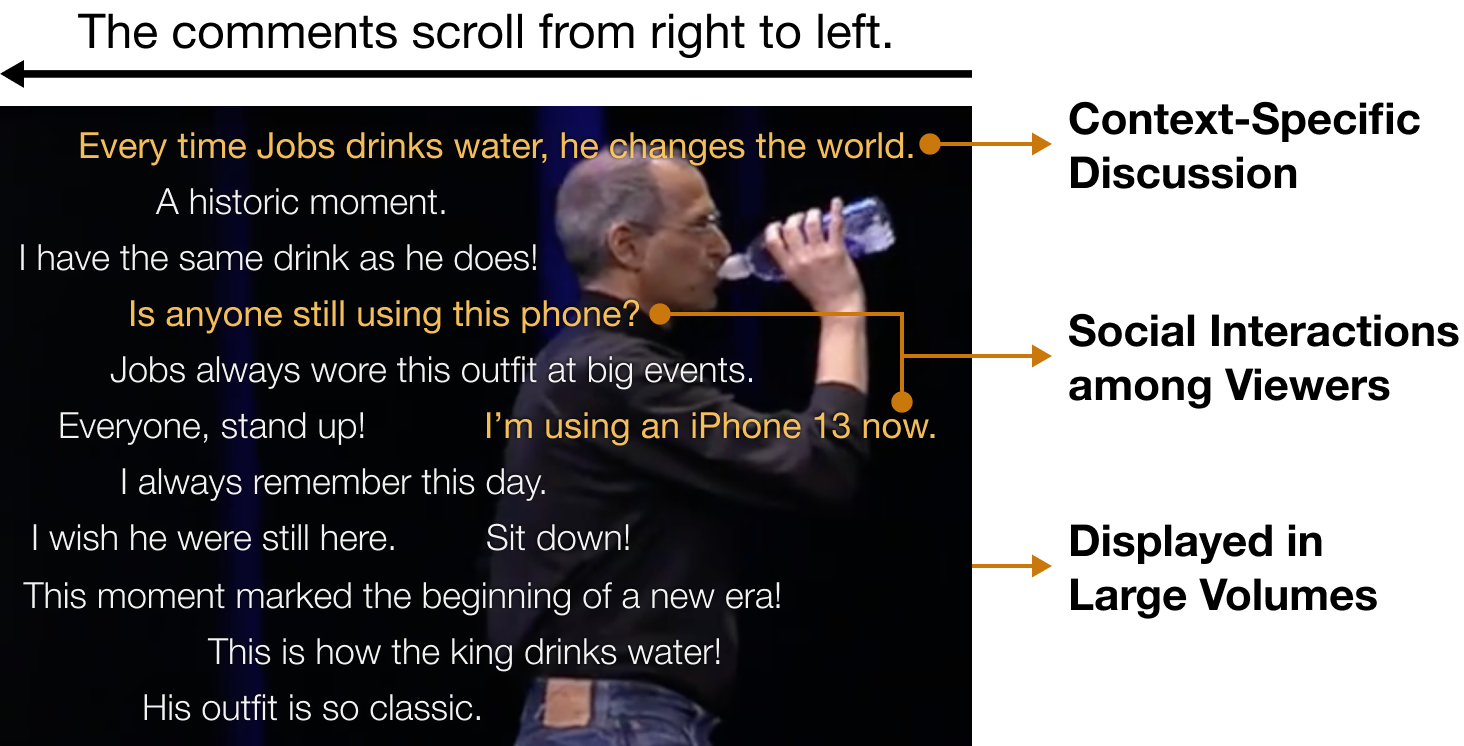}
    \caption{The key characteristics of Danmu comments.}
    \Description{This figure shows three key characteristics of Danmu comments. The first characteristic is that Danmu comments contain real-time, context-specific information about the video. The second characteristic is that Danmu comments involve social interactions among viewers. The third characteristic is that Danmu comments are displayed in large volumes simultaneously.}
    \label{fig:background}
\end{figure}

Compared to traditional comments, Danmu creates a closer connection between the video and the commentary \cite{ma2017video,chen2017watching}. 
Danmu comments are synchronized with the video timeline and visually overlaid on the screen, resulting in several unique features (see Figure~\ref{fig:background}). 
First, Danmu comments appear exactly at the video moment when they are posted. This allows viewers to exchange real-time, \textbf{context-specific} information (e.g., identifying an actor in a scene) rather than post-viewing reflections \cite{chen2017watching}. Second, Danmu comments are immediately displayed on the screen, visible to both the poster and other viewers. This enables more direct \textbf{social interaction among users}, creating a pseudo co-watching experience \cite{ma2017video,chen2017watching}. Third, Danmu comments often come in \textbf{large volumes} simultaneously \cite{lv2019understanding,wu2018danmaku}. This concurrent display enables sighted viewers to skim through the information efficiently. Fourth, Danmu comments are anonymous \cite{ma2017video,chen2022danmuvis}—usernames are hidden, and comments are not structured into threads. This anonymity leads viewers to address each other through semantic references \cite{ma2017video,he2017exploring}. For instance, users might reply with phrases like ``\textit{The one saying `Why not buy an airplane', are you serious?}'' or repeat the same comment to express shared opinions.

The HCI community has emphasized Danmu's role in enhancing social connection: it creates a co-watching experience \cite{ma2017video,chen2017watching,wu2019danmaku}, facilitates information exchange \cite{chi15learningDanmu,he2021beyond,chen2024towards}, and fosters a sense of belonging \cite{xiang2022influence,ma2017video}. However, the lack of accessibility support for Danmu may prevent certain individuals from participating in this social interaction. 
For example, people who are hard of hearing may find it challenging to engage with text-based Danmu comments \cite{chen2024towards}, as they often prefer sign language over text for information access. More significantly, Danmu's inherently visual design presents major challenges for BLV viewers. Despite these barriers, there has been limited understanding on how BLV viewers engage with Danmu. To fill this gap, we conducted a study where we invited BLV viewers to participate in co-watching sessions. Our work identified three key challenges that hinder BLV viewers' engagement with Danmu: the lack of visual context, the speech interference between comments and videos, and the disorganization of comments. In response, we designed DanmuA11y to address these challenges and create a socially engaging video viewing experience for BLV viewers.

\subsection{Social Media Accessibility}
Addressing the accessibility of social media has been a consistent focus in HCI research \cite{gleason2019addressing,gleason2020future}. The emergence of new types of visual content continues to challenge BLV users in fully participating on social media platforms \cite{voykinska2016blind,seo2022challenges,seo2021understanding}. In response, researchers have explored methods to help BLV users access various forms of visual content, including images \cite{wu2017automatic, gleason2020twitter,10.1145/3654777.3676336}, memes \cite{gleason2019making}, GIFs \cite{gleason2020making,zhang2022ga11y}, emojis \cite{tigwell2020emoji,zhang2021voicemoji}, comics \cite{huh2022cocomix}, and videos \cite{liu2021makes,wang2021toward,van2024making}. When engaging with social media content, BLV users seek more than just factual descriptions (e.g., identifying objects in an image); they also aim to establish social and emotional connections with other users \cite{gleason2019making}. For instance, BLV users hope to understand the humor in memes \cite{gleason2019making}, interpret comics through others' perspectives \cite{huh2022cocomix}, and share emotional resonance with other video viewers \cite{jiang2024s,caro2016testing}.

To help BLV viewers engage with others' perspectives, providing accessible user comments play a crucial role \cite{wang2021revamp,huh2022cocomix}. Tools like Cocomix \cite{huh2022cocomix} utilizes user comments to offer additional context to comics, enabling BLV users to enjoy humor and engage with the community. 
However, unlike traditional user comments, Danmu comments are time-synced and visually overlaid onto videos, presenting unique challenges for BLV users. Our research addresses these challenges by integrating Danmu comments into videos as multi-viewer audio discussions, enabling BLV users to engage more deeply with other audiences. 
By studying time-synced video comments, our research offers implications for broader contexts, such as enhancing the accessibility of real-time commentary on live-streaming platforms \cite{rong2022feels}.

\subsection{\hl{Video Accessibility for BLV Viewers}}
\hl{To enhance video accessibility for BLV viewers, prior research has explored various methods, including audio descriptions and spatial audio representations. In this section, we review these techniques and discuss how our work extends existing literature.}

\subsubsection{Audio Descriptions}
Audio descriptions make videos accessible to BLV viewers by narrating visual elements in sync with the video content \cite{campos2020cinead}. Prior works have summarized guidelines for creating effective audio descriptions \cite{pavel2020rescribe}: (1) Describe important visual elements that are essential for understanding the video, (2) Do not overlap audio descriptions with the video's dialogue, and (3) Use an objective style and avoid subjective interpretation. \hl{Unlike audio descriptions that provide objective visual information, Danmu comments emphasize social discussions among viewers and often contain subjective opinions (e.g., ``\textit{His outfit is so classic.}'') and emotional reactions (e.g., ``\textit{A historic moment!}'') \cite{ma2017video}, leading to distinct information needs. In response, our work identifies BLV viewers' needs for Danmu access through interviews and co-watching exercises with end users.}

Prior research has explored various methods to facilitate audio description creation, including manual \cite{killough2023exploring}, semi-automatic \cite{liu2022crossa11y,pavel2020rescribe}, and automatic approaches \cite{wang2021toward}. For instance, CrossA11y \cite{liu2022crossa11y} offers semi-automatic support for authors by identifying accessibility issues in videos through the detection of visual-auditory inconsistencies. Similarly, Rescribe \cite{pavel2020rescribe} utilizes dynamic programming to optimize the placement of audio descriptions within videos. Tiresias \cite{wang2021toward} presents a fully automatic pipeline that detects visual-auditory discrepancies, generates audio descriptions, and refines them into a coherent output. \hl{CustomAD \cite{natalie2024audio} further enables BLV viewers to personalize the content and style of audio descriptions. Compared to audio descriptions, Danmu comments are abundant and disorganized, presenting new challenges in content curation and integration. Our work addresses these challenges by proposing an automatic pipeline for curating and integrating Danmu comments, ensuring a seamless video viewing experience.}

\subsubsection{Spatial Audio Representations.}
Besides audio descriptions, prior works \cite{chang2022omniscribe,uist2023front_row,ning2024spica,dang2024musical} have explored using spatial audio to represent spatial information in videos. For instance, Front Row \cite{uist2023front_row} employs spatial audio to encode the positions and movements of players in sports broadcasts, allowing BLV viewers to directly perceive the action in real time. Similarly, SPICA \cite{ning2024spica} leverages spatial sound to help BLV viewers identify the location of objects within video key frames. Following this, \hl{Dang et al. \cite{dang2024musical} utilizes spatial audio descriptions to enhance BLV viewers' spatial understanding of music performances in virtual reality. 
Recent research \cite{immohr2023proof,immohr2024subjective} has demonstrated that spatial audio can enhance social presence in multi-person remote communication. Building on this insight, our work explores transforming Danmu comments into multi-viewer spatial audio discussions to enhance BLV users' social presence. Evaluation results showed that this approach effectively strengthened BLV viewers' sense of co-watching videos with others. Based on our findings, we derive insights for improving BLV users' social engagement in broader contexts.}

%% file: Sections/3_FormativeStudy.tex
\section{Formative Study}
We conducted a formative study with eight BLV viewers to understand their needs, practices, and challenges in accessing Danmu. 
\hl{Through interviews and co-watching exercises, we addressed the following questions: }

\hl{(1) \textbf{Motivations}: Why do BLV viewers want to access Danmu?}

\hl{(2) \textbf{Practices}: How do BLV viewers currently access Danmu? }

\hl{(3) \textbf{Challenges}: What challenges do BLV viewers encounter in accessing Danmu?}

\subsection{Methods}
\subsubsection{\textbf{Participants}}
We recruited eight BLV viewers (P1-P8; four male, four female; Table~\ref{tab:demographics} lists their demographics) who frequently watched videos on Danmu-enabled platforms. These participants were recruited from an online support community, with ages ranging from 26 to 38 (mean = 31.2, SD = 4.4). Four participants were totally blind and four participants were legally blind with light perception. All participants had prior experience accessing Danmu on platforms such as Bilibili, Douyin\footnote{https://www.douyin.com/}, and Youku\footnote{https://www.youku.tv/?lang=en\_US}. They mainly used these platforms on mobile phones and their viewing frequency ranged from several times a week to daily. All participants were native Mandarin speakers.

\subsubsection{\textbf{Design probe}}

Prior to the study, seven out of eight participants identified an mobile application (Etong)\footnote{https://apps.apple.com/app/id6471562648} as the \textbf{best currently available tool} for accessing Danmu. This tool presents Danmu comments in an auto-scrolling list but is limited to live-streaming. 
Since no Danmu-enabled video platforms offer similar functionality, we replicated Etong's interface in our design probe.

\begin{figure}[h]
    \centering
    \includegraphics[width=0.49\textwidth]{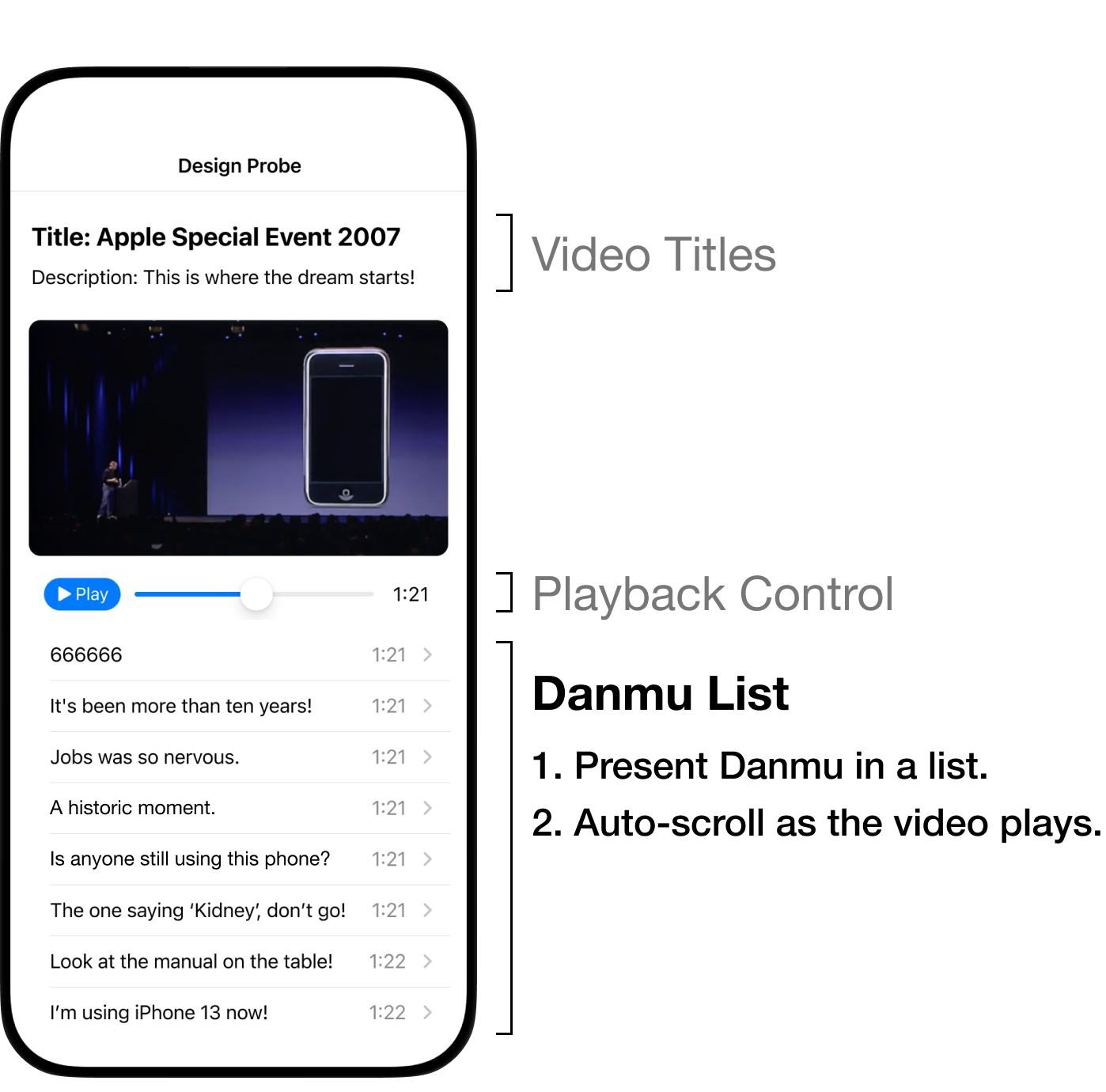}
    \caption{The design probe presents Danmu comments in an auto-scrolling list. Participants used the probe in Mandarin.}
    \Description{This figure shows the user interface of the design probe. The probe presents Danmu comments in an auto-scrolling list, which is accessible via screen readers.}
    \label{fig:design_probe}
\end{figure}

As shown in Figure~\ref{fig:design_probe}, the design probe displays Danmu comments in a list, which is accessible via screen readers \cite{jain2021smartphone,huh2023genassist}. The list auto-scrolls as the video plays, allowing users to easily view the latest comments. Additionally, the probe supports standard video playback controls, including a play/pause button and a slider.

The design probe contained six videos, each representing a genre commonly watched by BLV viewers \cite{jiang2024s}: an educational video, a comedic video, a tutorial video, a news video, a music video, and a film clip. The video lengths ranged from 33 seconds to 7 minutes (mean = 3.5 minutes, SD = 2.2 minutes). Table~\ref{tab:video_formative} lists their details. These videos were randomly chosen from videos with over 1,000 Danmu comments on Bilibili. 
We downloaded the latest 1,000 Danmu comments, the default maximum provided by the platform. 
We imported videos in MP4 format, Danmu comments in XML format, and creator-written titles and subtitles as texts into the design probe. The probe was implemented as an iOS app on an iPhone 15 and has been tested for compatibility with VoiceOver\footnote{VoiceOver is the built-in screen reader on iOS devices.}.

\subsubsection{\textbf{Procedure}}
The study consisted of two successive phases: a 45-minute interview followed by a 45-minute co-watching exercise, with a 5-minute break in between.

\textbf{Phase 1: Semi-structured Interview.} We asked participants about their motivations, practices, and challenges in accessing Danmu. To elicit details, we asked participants to show their practices on their preferred platforms. When interesting points arose, we followed up with questions to probe further details.

\textbf{Phase 2: Co-watching Exercise.}
In this session, participants first received a brief tutorial on the design probe and then used it to view the six pre-selected videos in a random order. They were encouraged to think aloud about any confusion while viewing videos. After the session, we conducted an exit interview to understand the challenges of using the design probe. The entire study was conducted one-on-one, in-person. Participants were compensated approximately 17 USD in local currency for their time.

\subsubsection{\textbf{Analysis}}
We recorded the study audio, transcribed it\footnote{Participants’ original speech was translated from Mandarin to English.}, and then used an open-coding approach \cite{corbin2015basics} to analyze participants' motivations, current practices, and challenges in accessing Danmu. Two authors independently reviewed the transcripts, developing and applying codes through an iterative process until reaching a consensus. The codes were then grouped into clusters representing the emerging themes from the study data. After a thorough discussion and review of all potential themes, the final themes were reported as the study findings.

\subsection{Findings}
\subsubsection{\textbf{Motivations for Accessing Danmu}}\label{sec:motivations}
Participants primarily viewed Danmu as a \textit{real-time} communication channel, where they sought comments that are ``\textit{closely related to the current video moment}'' (P1). During the interview, they mainly reported three motivations for accessing Danmu:

\textbf{(1) To Enjoy Creative Comments}: Seven participants sought joy from Danmu comments that were ``\textit{creative}'' (P1), ``\textit{unexpected}'' (P3), or ``\textit{hilarious}'' (P7). For example, P2 found a comment in a bakery video particularly creative: ``\textit{Use soy sauce if you don't have dark chocolate.}'' Similarly, P5 described some Danmu comments as ``\textit{non-visual bloopers}'', citing the unexpected comment, ``\textit{There is a bottle of mineral water on the table!}'' in a historical drama. These comments often added an element of surprise and enhanced the enjoyment of their viewing experience.

\textbf{(2) To Gain Supplementary Information}: Six participants valued Danmu for providing additional information not present in the original video. For instance, P5 recalled many informative Danmu comments when Steve Jobs released the first iPhone, such as ``\textit{Jobs was really nervous because there were so many issues with this prototype iPhone.}'' Such information ``\textit{enriched the video}'' (P5) and ``\textit{deepened their understanding of the moment}'' (P8).

\textbf{(3) To Engage with Diverse Opinions}: Six participants appreciated the diverse perspectives in Danmu comments. For example, P2 noted that in a video discussing whether to buy a smart speaker, viewers shared contrasting opinions like, ``\textit{Don't buy it! Its sound quality is bad.}'' and ``\textit{You can't miss it! The voice assistant is really useful.}'' Such debates ``\textit{offered new perspectives}'' (P5), ``\textit{heated up the discussion}'' (P7), and ``\textit{made the conversation more engaging}'' (P2).

\subsubsection{\textbf{Current Practices}}
Despite the benefits of Danmu, all participants noted that accessibility support for Danmu on mainstream video platforms is quite limited. These platforms either offer only a single Danmu comment (near the playback time) or make all comments inaccessible to screen reader users, forcing participants to adopt resourceful \textbf{workarounds}.

Seven participants attempted to use screen recognition tools (e.g., VoiceOver Recognition\footnote{https://support.apple.com/en-us/111799}) to access Danmu as on-screen text but found them ineffective due to issues like ``\textit{having to pause the video}'' (P2) and ``\textit{recognition errors from cluttered text}'' (P5). In addition to technical workarounds, three participants mentioned asking friends to read out interesting Danmu comments when watching videos together, while two occasionally encountered videos that curated Danmu comments in other videos. However, all participants agreed that ``\textit{the opportunity to get human support is limited}'' (P6). Overall, the lack of accessible support for Danmu excludes participants from fully enjoying Danmu and engaging in the social discussions.

\subsubsection{\textbf{Three Key Challenges}}
During the co-watching exercise, participants primarily read Danmu while watching videos because ``\textit{it feels like a live discussion happening alongside the video}'' (P3). All participants described the design probe as ``\textit{better than any existing tools}'' (P1) because ``\textit{it allows us to access what is otherwise mostly hidden}'' (P7). However, they still faced three significant challenges in fully engaging with Danmu:

\textbf{First, Danmu comments often lack descriptions of the visual context, making it difficult for BLV viewers to fully understand the discussions.} For example, four participants were confused by the comment, ``\textit{Is she living in summer or winter?}'', which humorously referred to an actress wearing both a winter scarf and a summer T-shirt. However, BLV viewers found it confusing because the visual context was inaccessible. This lack of visual context hindered their ability to grasp the topics and ``\textit{achieve the same level of understanding as sighted viewers}'' (P1).

\textbf{Second, the speech interference between Danmu comments and videos prevents BLV viewers from enjoying both content simultaneously.} Although participants wanted to access Danmu while watching videos, the comments often overlapped with the original speech, making it ``\textit{difficult to grasp both at once}'' (P2). Furthermore, Danmu comments often appeared incoherent with the surrounding video speech. For example, P1 noted, ``\textit{Many comments seemed irrelevant to the original video, making them more like a noise.}'' This lack of coherence resulted in an experience that was ``\textit{obtrusive}'' (P4) and ``\textit{distracting}'' (P7).

\textbf{Third, Danmu comments are not structured for sequential access, making it tedious for BLV viewers to follow audience discussions.} Danmu comments typically come in large volumes and appear on screen simultaneously. However, when presented sequentially by screen readers, they become disorganized and shift rapidly between topics (e.g., ``\textit{Any one still using this phone? ; The one saying `Kidney', don't go! ; Look at the manual on the table! ; ...}''). This disorganization of comments made it challenging for participants to follow audience discussions, leading to an experience that felt ``\textit{more like a reading test rather than an enjoyable chit-chat}'' (P8), ultimately preventing them from connecting with other viewers.

Overall, while Danmu serves as a valuable source of social engagement, participants encountered significant challenges due to the lack of visual context, the speech interference between comments and videos, and the disorganization of comments. To address these challenges, we designed DanmuA11y.

%% file: Sections/4_System.tex
\FloatBarrier

\begin{figure*}[ht!]
    \centering
    \includegraphics[width=\textwidth]{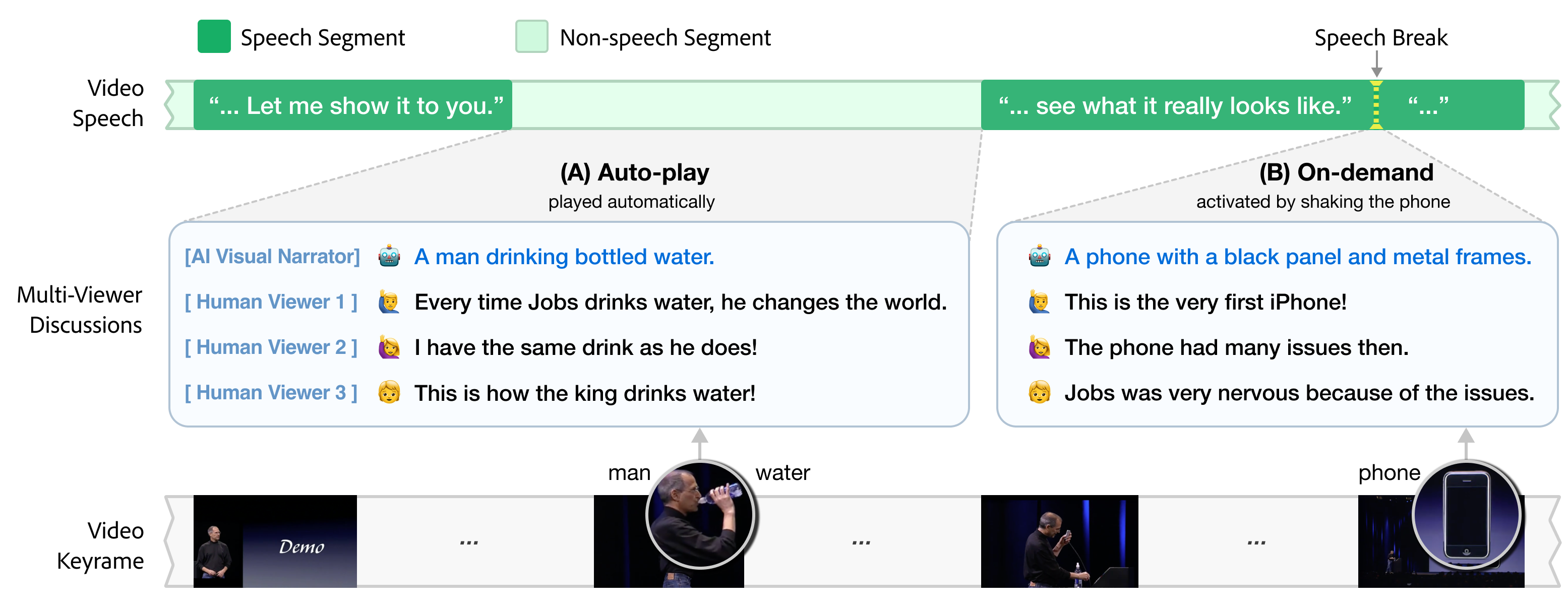}
    \caption{An example walk-through of DanmuA11y: (A) The system \textbf{automatically plays} multi-viewer discussions during non-speech segments, using spatial audio and varied tones to create a co-watching experience. (B) The system provides \textbf{on-demand access} at speech breaks, notifying users with a sound and allowing them to access Danmu comments by shaking their phones.}
    \Description{This figure shows an example walk-through of DanmuA11y. It contains two examples. In the first example, the system automatically plays multi-viewer discussions during non-speech segments, using spatial audio and varied tones to create a co-watching experience. In the second example, the system provides on-demand access to multi-viewer discussions at speech breaks, notifying users with a sound and allowing them to access Danmu comments by simply shaking their phones.}
    \label{fig:walk-through}
\end{figure*}

\section{System}

DanmuA11y makes Danmu accessible by transforming Danmu into multi-viewer discussions. It incorporates three core features: 

(1) \textit{\textbf{Augmenting Danmu with Visual Context}}: To help BLV viewers understand the visual context related to audience discussions, DanmuA11y supplements Danmu comments with descriptions of the visual context. 

(2) \textit{\textbf{Seamlessly Integrating Danmu into Videos}}: To help viewers enjoy Danmu while watching videos, DanmuA11y curates Danmu topics and optimizes their insertion timing within videos to avoid Danmu-speech overlaps. 

(3) \textit{\textbf{Presenting Danmu via Multi-Viewer Discussions}}: To create a co-watching experience, DanmuA11y organizes Danmu comments into dialogues and uses spatial audio to simulate the sensation of other viewers conversing around the user.

Powered by these features, DanmuA11y offers an enjoyable and socially engaging experience for BLV viewers. In the following sections, we first provide an example walk-through in Section~\ref{sec:walkthrough}. Next, we describe the system pipeline in Sections \ref{sec:pipeline_overview} - \ref{sec:system_sub4}. Finally, we report the implementation details in Section~\ref{sec:implementation}.

\subsection{User Walk-through}\label{sec:walkthrough}

To demonstrate DanmuA11y, we follow Jason, who is watching a video of Steve Jobs launching the first iPhone (see Figure~\ref{fig:walk-through}). He uses DanmuA11y to engage with the audience discussions in Danmu comments.

DanmuA11y \textbf{seamlessly integrates} Danmu comments with the video, ensuring they do not overlap with the video's speech. For instance, during Steve Jobs' speech, the system does not insert Danmu comments, allowing Jason to focus on the speech. When the system detects a non-speech segment in the video, such as after Steve says, ``\textit{... Let me show it to you}'', it \textbf{automatically plays} a sequence of curated Danmu comments: ``[Visual Description] \textit{A man is drinking bottled water.} [Viewer 1] \textit{Every time Jobs drinks water, he changes the world.} [Viewer 2] \textit{I have the same drink as he does!} ...'' (see Figure~\ref{fig:walk-through} (A)). These comments are curated for their creativity and informativeness, organized to reflect interactions among multiple viewers, and preceded by a visual description to help Jason grasp the visual context. Hearing these comments, Jason feels like he is co-watching with other viewers, because the comments are delivered using spatial audio and varied human tones, creating the sensation of live discussions happening around him.

The system also provides \textbf{on-demand Danmu access} at speech breaks. To minimize disruption, the system plays a lightweight notification sound at detected speech breaks. For example, after Steve says, ``\textit{... You get to see what it really looks like}'', Jason hears a notification sound and becomes curious about what others are discussing. To access these discussions, Jason simply shakes his phone to pause the video and listen to the comments: ``[Visual Description] \textit{A phone with a black front panel and metal frames.} [Viewer 1] \textit{This is the very first iPhone!} [Viewer 2] \textit{The phone had many issues then} ...'' (see Figure~\ref{fig:walk-through} (B)). These comments help Jason gain a deeper understanding of this video moment. Once the comments conclude, the video auto-rewinds to the previous speech break and resumes at the start of a sentence, allowing Jason to refocus on the video easily.

\subsection{Pipeline Overview}\label{sec:pipeline_overview}

\begin{figure*}[ht!]
    \centering
    \includegraphics[width=\textwidth]{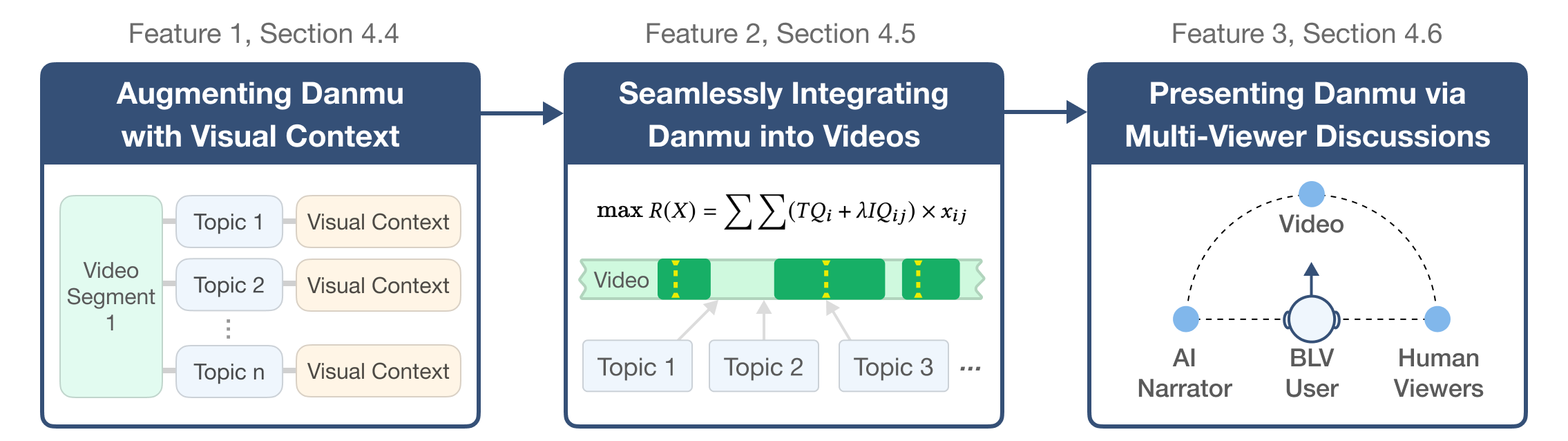}
    \caption{The pipeline of DanmuA11y.}
    \label{fig:pipeline}
    \Description{This figure shows the pipeline of DanmuA11y. It contains three features. The first feature is Augmenting Danmu with Visual Context. The second feature is Seamlessly Integrating Danmu into Videos. The third feature is Presenting Danmu via Multi-Viewer Discussions.}
\end{figure*}

As shown in Figure~\ref{fig:pipeline}, DanmuA11y incorporates three features. 
The first feature supplements Danmu topics with descriptions of the visual context. The second feature integrates Danmu topics into the video, employing an optimization method to maximize both topic quality and insertion quality. The third feature transforms Danmu topics into multi-viewer audio discussions. 
In the following, we first describe the video segmentation method that underpins these features, followed by a detailed introduction to each feature.

\FloatBarrier

\subsection{Video Segmentation}\label{sec:system_sub1}

\begin{figure*}[ht!]
    \centering
    \includegraphics[width=0.75\textwidth]{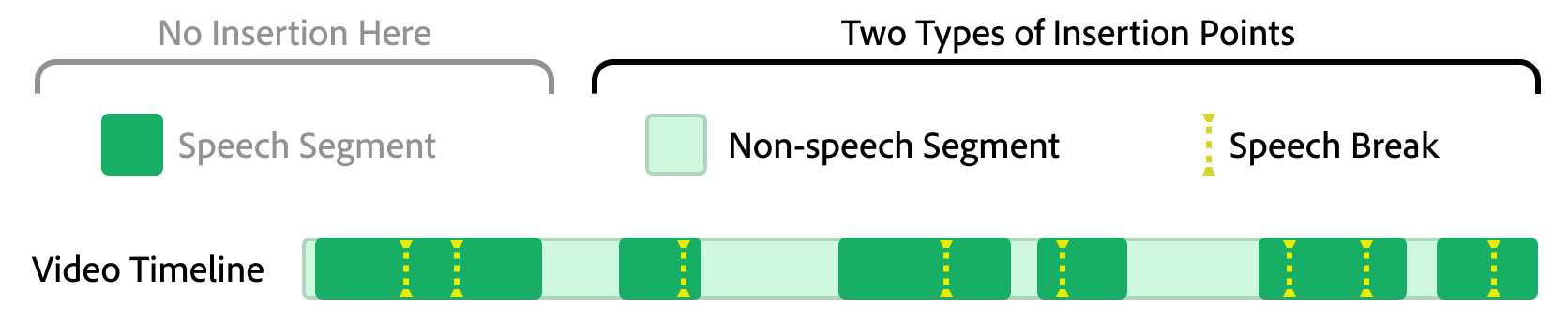}
    \caption{DanmuA11y divides the video into non-speech and speech segments. The points between adjacent speech segments are called speech breaks.}
    \label{fig:module1}
    \Description{This figure shows the video segments on a video timeline. The video timeline is divided into non-speech segments and speech segments. The points between adjacent speech segments are identified as speech breaks.}
\end{figure*}

DanmuA11y divides the video into two types of segments: non-speech and speech segments. This speech-based segmentation allows the system to identify proper insertion points for Danmu.

\subsubsection{\textbf{Non-Speech Segment}}
A non-speech segment is a continuous period free of human speech, providing a space for Danmu insertions without overlapping with the spoken content. These segments are identified using time-aligned transcripts generated by the Volcengine Auto-Subtitle API\footnote{https://www.volcengine.com/}, which provides start and end times for each sentence with millisecond-level precision (e.g., ``\textit{00:00:01,010 --> 00:00:04,100 Hello, welcome to the video!}''). Following prior research \cite{liu2022crossa11y}, we classify any gap between sentences that lasts longer than two seconds as a non-speech segment. Furthermore, to prevent the insertion of Danmu in non-speech areas with high volumes (e.g., loud background music), we apply a one-second sliding window to identify regions where the root-mean-square volume exceeds a threshold of 0.8 using librosa\footnote{librosa.org}. These high-volume areas are marked as inappropriate for insertion.

\subsubsection{\textbf{Speech Segment}}
After identifying non-speech segments, we obtain the initial speech segments. These speech segments are further divided into smaller segments by speech breaks. A speech break is a specific point that separates continuous speech into relatively independent paragraphs. This allows the video to be paused without significantly disrupting the original speech flow. We identify these breaks using the natural language processing capabilities of GPT-4o \cite{achiam2023gpt}, inputting consecutive sentences (i.e., without any non-speech segments in between) and prompting the model with ``\textit{Split this text into several segments, ensuring that each segment is relatively independent.}'' This process typically identifies break points where the topic or idea changes (e.g., ``\textit{... That's our new product.} [Break point] \textit{Now let's move on to the demo...}''). To avoid excessive pauses, each speech segment is constrained to have at least 20 words. The end of each speech segment, unless it is followed by a non-speech segment, is then marked as a speech break. The identified video segments and speech breaks are subsequently utilized by the three core modules.

\subsection{Feature 1: Augmenting Danmu with Visual Context}\label{sec:system_sub2}
Danmu comments often discuss specific visual elements that are inaccessible to BLV viewers, making it challenging for them to understand the discussion topics. To address this issue, DanmuA11y groups the comments into topics and supplements these topics with descriptions of the visual context.

\begin{figure*}[h]
    \centering
    \includegraphics[width=0.75\textwidth]{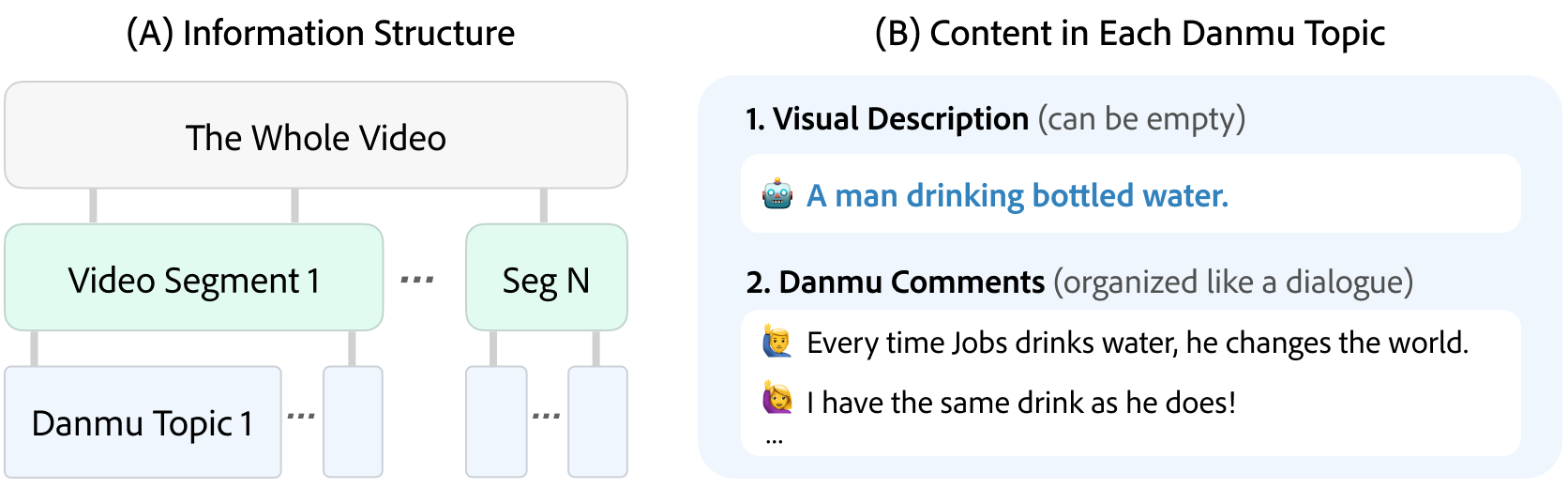}
    \caption{(A) Danmu topics belong to each video segment. (B) Each topic includes a visual description and a list of comments.}
    \label{fig:module2}
    \Description{This figure shows the example of a Danmu topic. Danmu topics belong to each video segment. Each topic includes a visual description and a list of Danmu comments. In this example, the visual description is "A man drinking bottled water." The list of Danmu comments contain the following items: [Comment 1]"Every time Jobs drinks water, he changes the world.", [Comment 2] "I have the same drink as he does!"}
\end{figure*}

\subsubsection{\textbf{Grouping Danmu into Topics}}
To identify the main topics in each video segment, we utilize a prompt-based topic modeling method \cite{pham2023topicgpt}. Specifically, we feed GPT-4o all the Danmu comments\footnote{We use the video time associated with Danmu comments to assign them to their respective segments.} from a segment and prompt\footnote{The full prompts are provided in the supplementary materials.} it to ``\textit{Summarize the main topics discussed by viewers and categorize the comments under each topic.}'' This approach generates a list of topics, each accompanied by a brief summary and their associated comments. Subsequently, each topic is re-input into GPT-4o with instructions to ``\textit{Remove redundant comments and rearrange the remaining ones to simulate a dialogue among multiple viewers.}'' After this step, each video segment contains several topics, with each topic including a reordered list of comments that simulate a conversation among viewers.

\subsubsection{\textbf{Generating Visual Descriptions}}
To help BLV viewers understand the discussion topics, DanmuA11y generates visual descriptions for topics using GPT-4o \cite{hu2023promptcap,achiam2023gpt}. Specifically, the GPT-4o model is provided with the topic summaries and key frames for each video segment, and is prompted with the following instruction: ``\textit{Determine whether each topic discusses a visual object in the key frames. If so, describe the object's appearance in 15 words or fewer. If not, output `None'.}'' The key frames are pre-extracted from the video using SceneDetect\footnote{https://www.scenedetect.com/}, a content-aware scene detection model that divides the video into shots by comparing adjacent frames in the HSV color space. For each shot, the middle frame is selected as the representative key frame. These key frames are assigned to their respective video segments based on their timestamps. After this step, each Danmu topic includes a visual description (which may be empty) and a list of Danmu comments. Figure~\ref{fig:module2} (B) illustrates an example of the content within a topic.

\subsection{Feature 2: Seamlessly Integrating Danmu into Videos}\label{sec:system_sub3}
To ensure viewers can enjoy Danmu while watching videos, DanmuA11y curates Danmu topics and determines their insertion points using an optimization algorithm. \hl{The algorithm optimizes two aspects: (1) \textbf{Topic Quality}: the quality of Danmu topics, and (2) \textbf{Insertion Quality}: the suitability of inserting Danmu topics at specific points in the video.}

\begin{figure*}[h]
    \centering
    \includegraphics[width=0.8\textwidth]{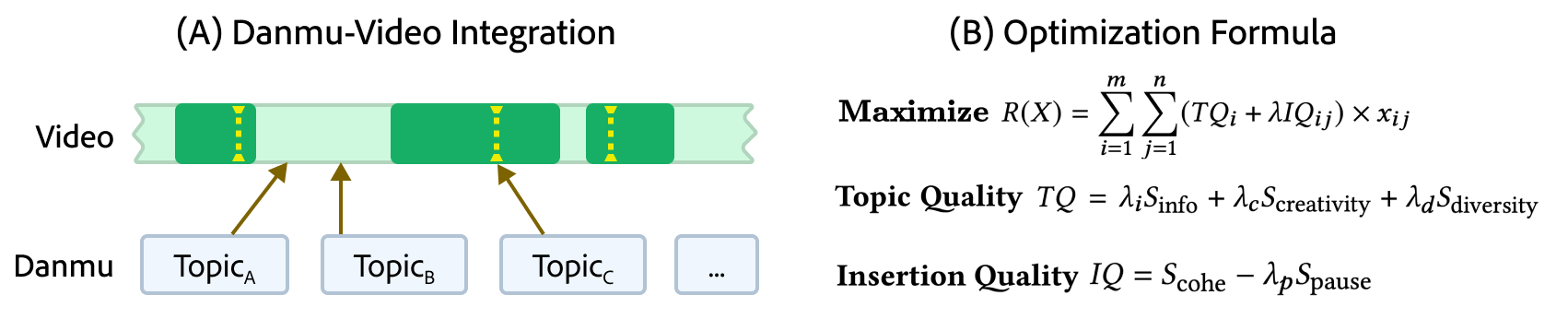}
    \caption{(A) The system places Danmu topics into insertion points. (B) It maximizes both topic quality and insertion quality.}
    \label{fig:module3}
    \Description{This figure shows the Danmu-video intergation method. The system places Danmu topics into insertion points in the video. To determine the integration, the system maximizes both topic quality and insertion quality.}
\end{figure*}

\subsubsection{\textbf{Topic Quality}}
DanmuA11y curates Danmu topics based on the three metrics derived from the formative study (as shown in Section~\ref{sec:motivations}): informativeness, creativity, and opinion diversity. Each metric is measured using established methods from prior works \cite{park2023generative, wang2021toward, liu2023coargue}.

\textbf{Informativeness} assesses how much new information a topic adds to the original speech. Following \cite{wang2021toward}, we quantify informativeness by measuring the semantic correlation between a Danmu topic and the transcript of the corresponding video segment. Specifically, we generate their vector embeddings using the \textit{Universal Sentence Encoder} \cite{cer2018universal} and compute the cosine similarity $sim$ between the two encoded vectors. The informativeness score $S_{\text{info}}$ is then defined as $1-sim$, where a low semantic correlation results in a high informativeness score.

\textbf{Creativity} measures how creative a topic is. Given the subjective nature of creativity, we follow \cite{park2023generative} by using GPT-4o for subjective ratings. The model is provided with the topic's content and instructed to ``\textit{Rate the creativity on a scale of 1 to 10, where 1 indicates completely mundane and 10 indicates exceptionally creative.}'' This results in a rating of 8 for unusual descriptions like ``\textit{The snack looks like feet.}'' and a rating of 2 for greetings like ``\textit{I'm here for a second time!}''. These ratings are then normalized to a 0-1 scale by dividing by 10, yielding the creativity score $S_{\text{creativity}}$.

\textbf{Opinion Diversity} measures the number of different opinions within a topic, as our formative study shows that topics with different opinions could provide a sense of heated discussion. To quantify opinion diversity \cite{liu2023coargue}, we apply sentiment analysis using the \textit{XLM-T}\footnote{https://huggingface.co/cardiffnlp/twitter-xlm-roberta-base-sentiment} model \cite{barbieri2021xlm}, which classifies each comment as ``\textit{positive}'', ``\textit{neutral}'', or ``\textit{negative}''. The diversity score $S_{\text{diversity}}$ is calculated by dividing the number of distinct sentiment labels by the total number of possible labels (i.e., three). A score of 1.0 indicates all three sentiments are present (e.g., ``[positive] \textit{My favorite snack;} [negative] \textit{It is too sour;} [neutral] \textit{I may give it a try; ...}'').

\textbf{Topic Quality} $\boldsymbol{TQ}$ is defined as the weighted sum of the three metrics: $TQ = \lambda_{i} S_{\text{info}} + \lambda_{c} S_{\text{creativity}} + \lambda_{d} S_{\text{diversity}}$. Each metric is scaled from 0 to 1. We set $\lambda_{i}=\lambda_{c}=\lambda_{d}=1$ to assign equal weight to each metric. The weights could be further personalized based on user preferences, which we discuss in Section~\ref{sec:personalization}.

\subsubsection{\textbf{Insertion Quality}}
Since Danmu topics are often related to nearby video moments, we restrict each topic's candidate set of insertion points to two locations before and two locations after the topic, including both non-speech segments and speech breaks. The quality of inserting a Danmu topic at each candidate point is measured based on two metrics: coherence and pause.

\textbf{Coherence.} Ideally, the inserted Danmu topic should be coherent with the original speech. We measure coherence using the method from \cite{pavel2020rescribe}, employing a pre-trained language model, \textit{davinci-002}\footnote{\textit{davinci-002} is a GPT base model. https://platform.openai.com/docs/models/gpt-base}, to estimate the likelihood of word sequences. We combine the inserted topic $t_i$ with the surrounding speech at the insertion point $p_j$, input the combined text into the model, and calculate the sum log probability of $t_i$ at $p_j$. We apply min-max normalization to the log probabilities for each topic, and use the resulting normalized values as the coherence score $S_{\text{cohe}}$.

\textbf{Pause Penalty.} Long pauses at speech breaks can distract users from the original video. To minimize unnecessary pauses, we apply a penalty based on pause duration. The duration is estimated by dividing the word count in a topic by three, representing a typical speech rate of three Chinese words per second. This value is then normalized by the maximum pause duration at speech breaks (set to ten seconds in this work), resulting in the pause score $S_{\text{pause}}$.

\textbf{Insertion Quality} $\boldsymbol{IQ}$ is defined as: $IQ = S_{\text{cohe}} - \lambda_{p} S_{\text{pause}}$. We set $\lambda_{p}=0.25$ to allow for pauses when necessary.

\subsubsection{\textbf{Insertion Optimization}}
When inserting Danmu topics into the video, our objective is to maximize both the topic quality \(TQ\) and the insertion quality \(IQ\), subject to the length constraints of insertion points. We formulate this as an optimization problem. We have a set of topics \(\{t_i\}\), where \(i = 1, \dots, m\), each having a length $l_i$, and a list of insertion points \(\{p_j\}\), where \(j = 1, \dots, n\), each having a maximum length $L_j$. Each topic has a topic quality score $TQ_i$ and there is an insertion quality score $IQ_{ij}$ for placing topic $t_i$ into point $p_j$. Our goal is to maximize the overall reward:

\[R(X) = \sum_{i=1}^{m} \sum_{j=1}^{n} (TQ_i + \lambda IQ_{ij}) \times x_{ij}\]

This optimization is achieved by determining the set of decision variables $X = \{x_{ij}\}$, where $x_{ij}$ represents whether topic $t_i$ is inserted into point $p_j$. The constraints are: (1) the total length of topics inserted into each point must not exceed the point's maximum length: $\sum_{i=1}^{m} l_i x_{ij} \leq L_j, \forall j = 1, 2, \ldots, n$; (2) each topic can be inserted at most once: $\sum_{j=1}^{n} x_{ij} \leq 1, \forall i = 1, 2, \ldots, m$; and (3) the decision variables $x_{ij}$ are binary: $x_{ij} \in \{0, 1\}, \forall i, j$. The optimal solution can be found by solving $X^* = \arg \max R(X)$ using integer linear programming. After obtaining the optimal solution $X^*$, we sort all the topics inserted at the same insertion point by their topic quality score in descending order.

In this work, the maximum length of each speech break is constrained to ten seconds to prevent overly long pauses. The length of each topic is dividing the word count by three, which reflects a typical speech rate of three Chinese words per second. The weight $\lambda$ is empirically set to 10 to prioritize insertion coherence.

Overall, this approach prioritizes inserting high-quality topics into non-speech segments first, followed by speech breaks, and ultimately discards low-quality topics if they do not fit within the available length constraints.

\subsection{Feature 3: Presenting Danmu via Multi-Viewer Discussions}\label{sec:system_sub4}
After inserting Danmu topics into the video, DanmuA11y transforms it into multi-viewer discussions using spatial audio and varied tones, aiming to create a socially engaging experience.

\begin{figure*}[h]
    \centering
    \includegraphics[width=0.75\textwidth]{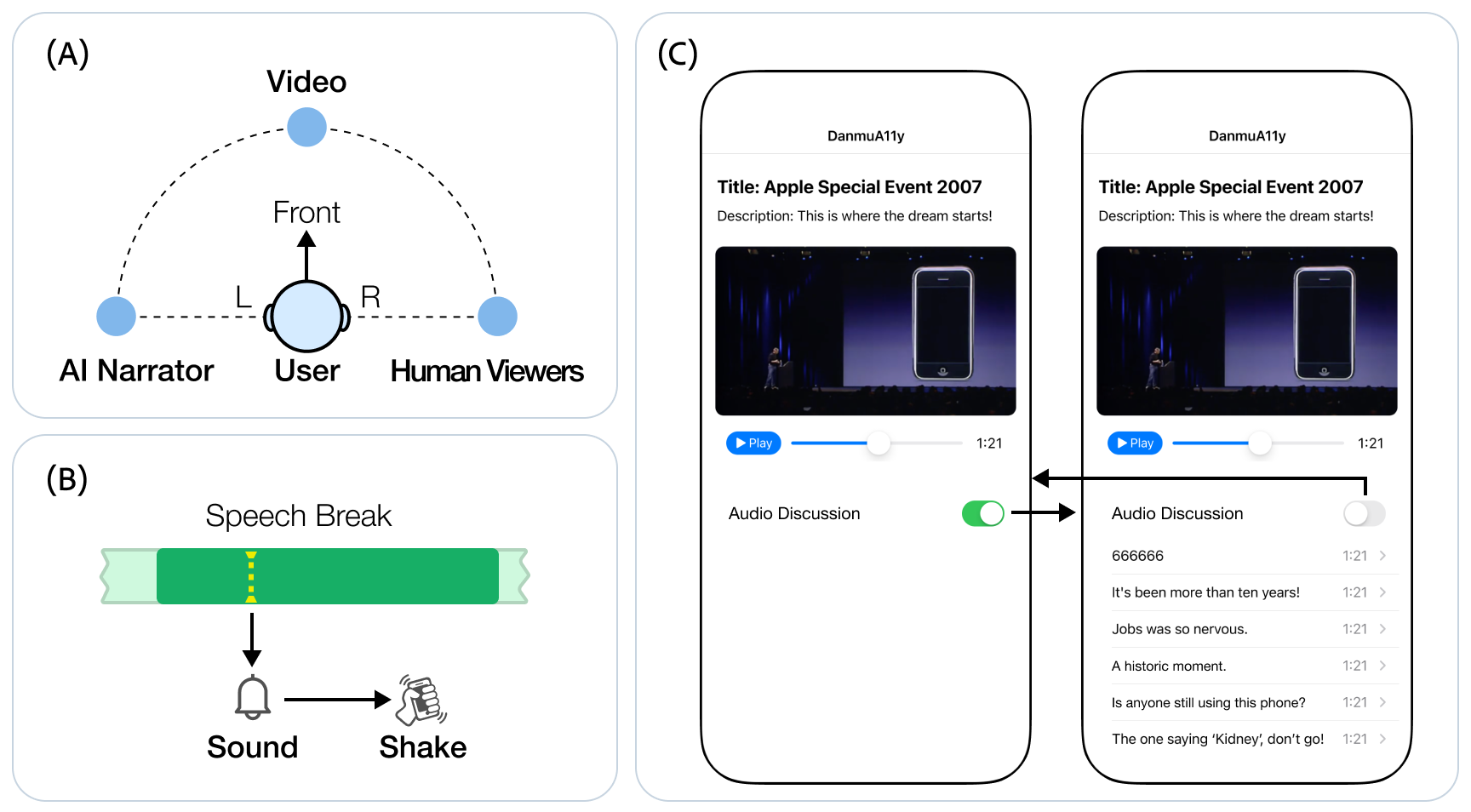}
    \caption{DanmuA11y presents Danmu via multi-viewer discussions. (A) The system positions the AI narrator on the left side and the human viewers on the right side, with the video placed in the front. (B) The system plays a notification sound at speech breaks, allowing users to shake their phones to access Danmu on demand. (C) The Audio-Discussion toggle is enabled by default, allowing users to engage with the multi-viewer discussions. If the toggle is turned off, the system displays the original Danmu list.}
    \label{fig:module4}
    \Description{This figure shows the multi-viewer discussion design of DanmuA11y. (A) The system positions the AI narrator on the left side and the human viewers on the right side, with the video placed in the front. (B) The system plays a notification sound at speech breaks, allowing users to shake their phones to access Danmu on demand. (C) The system contains an Audio-Discussion toggle. The toggle is enabled by default, allowing users to engage with the multi-viewer discussions. If the toggle is turned off, the system displays the original Danmu list.}
\end{figure*}

\subsubsection{\textbf{Multi-Viewer Audio Discussions}}
DanmuA11y transforms Danmu into audio discussions from two types of virtual viewers: (1) an \textit{AI narrator} who delivers visual descriptions and (2) multiple \textit{human viewers} who verbalize Danmu comments. These discussions are played via spatial audio, aiming to simulate people conversing around the user.

\textit{\textbf{Spatial Layout.}} As shown in Figure~\ref{fig:module4} (A), we position the AI narrator on the left side and the human viewers on the right side, with the video placed in the front. This setup ensures that each sound source is distinct from the others. All three sound sources are set to equal volume. To avoid concurrent speech, only one virtual viewer speaks at a time.

\textit{\textbf{Alternating Tones.}} We alternate the tones to simulate multi-person conversations. For each comment from human viewers, we randomly select a tone from three options and ensure that adjacent comments have different tones. A fourth tone is assigned to the AI narrator. The tones are synthesized using the Volcengine Text-to-Speech API\footnote{https://www.volcengine.com/product/tts. The tone IDs are BV002 for the AI narrator and BV011, BV064, and BV411 for human viewers.}.

\textit{\textbf{Auto-play Strategy.}} During non-speech segments, the audio discussions are auto-played alongside the video. At speech breaks, the system offers on-demand access to the discussions using the following design.

\subsubsection{\textbf{Easy On-demand Access}}\label{sec:on-demand-design}
When audio discussions are inserted at speech breaks, the system plays a notification sound to signal their presence, allowing users to access them selectively using a simple shake gesture.

\textit{\textbf{Notification Sound.}} We use a half-second bubble sound\footnote{The sound is provided in the supplementary materials.} to indicate available discussions. The sound plays from the left if the discussions include AI visual descriptions or from the right if they only contain viewer comments. The video continues to play (i.e., it is not paused) while the sound is played.

\textit{\textbf{Shake gesture.}} On hearing the notification sound, users can shake their phone within five seconds, which will pause the video and play the discussions. These discussions use the same spatial layout and tones as described earlier. When the comments end, the video auto-rewinds to the last speech break, resuming at the start of a sentence.

\textit{\textbf{Custom Controls.}} Besides the shake gesture, the system provides video playback controls, including a play/pause button and a slider. There is an Audio-Discussion toggle that enables or disables audio discussions (both auto-played and on-demand). The toggle is on by default, allowing users to enjoy Danmu alongside the video. Users can turn it off to watch only the original video, such as when focusing on instrumental music. When the toggle is turned off, the system shows the original Danmu list (see Figure~\ref{fig:module4} (C)), which is accessible via screen readers.

\subsection{Implementation}\label{sec:implementation}

\subsubsection{\textbf{System}} The system is implemented as an iOS app for iPhone, tested on iPhone 15 with AirPods Pro (second generation). The phone-shake gesture is triggered when the device acceleration exceeds $6 m/s^2$ (detected by CoreMotion API, excluding gravity). The audio playback is handled using AVFoundation.

\subsubsection{\textbf{Pipeline}} The video processing pipeline is implemented in Python. For modules utilizing GPT-4o, the temperature is set to 0.8. To solve the integer programming problem, we use the Gurobi Optimizer\footnote{https://www.gurobi.com} via a Python interface, which completes within five seconds for 50 topics and 30 insertion points. The audio discussions are rendered using the Volcengine Text-to-Speech API at a speech rate between 1.1 and 1.2, adjusted to fit the duration estimated by word count. Spatial audio is generated via Pydub and exported in MP3 format.

\subsubsection{\textbf{Video Dataset}}\label{sec:video_dataset}
We randomly selected 18 videos from Bilibili to represent six video types frequently watched by BLV viewers \cite{jiang2024s}: educational, comedic, tutorial, news, music, and film clips, with three videos for each type. \hl{None of the 18 videos were included in the formative study.} Table~\ref{tab:video_evaluation} provides the video details. All videos had over 1,000 Danmu comments. We downloaded the latest 1,000 Danmu comments, the default maximum provided by the platform. 
The selected videos ranged from 24 seconds to seven minutes in length (mean = 3.8 minutes, SD = 1.7 minutes). The speech ratio (i.e., ratio of speech segments) in each video varied from 0\% to 100\% (mean = 59\%, SD = 38\%).

\subsubsection{\textbf{Technical Performance}}

We ran the pipeline on the above 18-video dataset. None of the videos had been tested during the system's development. After processing the 18 videos, the pipeline yielded 706 Danmu topics. Based on the data, we report the technical performance regarding (1) visual description accuracy and (2) creativity scoring.

\textbf{(1) \textit{Visual Description Accuracy.}} To evaluate the visual description accuracy, we randomly sampled 200 topics with visual descriptions, reviewed the topics and their associated key frames, and labelled whether the visual descriptions matched the visual elements being discussed. One researcher labeled all the data, and a second researcher reviewed the labels to ensure reliability. \hl{The accuracy rate was calculated by dividing the number of correct descriptions by the total number of sampled descriptions, yielding an accuracy rate of 92.5\%. The errors mainly arose from difficulties in describing rapid visual changes. For example, in V11, a fast-paced human fight was mistakenly described as ``\textit{A person is performing CPR}'' based on the comment ``\textit{CPR}''. Additionally, incorrect associations between comments and visual elements were noted, such as linking the comment ``\textit{He's the CEO!}'' to the wrong person in the video. We discuss potential methods to mitigate these errors in Section~\ref{sec:generalizability}.}

\textbf{(2) \textit{Creativity Rating.}} To assess how well GPT-4o rated creativity, we analyzed the distribution of creativity scores across all 706 topics. The scores ranged from 0.1 to 0.8 ($\mu = 0.408,\ \sigma = 0.139$). Specifically, 32.1\% of the scores fell between 0.1 and 0.3, typically for greetings (e.g., ``\textit{I'm here!}'') or simple emotions (e.g., ``\textit{Haha! So funny.}''). The majority, 54.3\%, fell between 0.4 and 0.5, representing commonly seen comments (e.g., \textit{``This cake looks tasty.}''). The remaining 13.6\% scored between 0.6 and 0.8, mainly for unexpected descriptions (e.g., ``\textit{The snack looks like feet.}'') and bloopers (e.g., ``\textit{Wait! He didn't even pay the bill!}''). This distribution aligns with the general observation that creative comments are not commonly found. One issue identified was that the model assigned scores lower than five for certain memes or puns (e.g., using common words to denote a humorous fact), which could potentially be improved by fine-tuning the model with specialized datasets \cite{sun2023new}.

The performance of other modules can be found in prior works \cite{pham2023topicgpt, pavel2020rescribe, wang2021toward, liu2023coargue}. Additionally, we evaluated the overall Danmu curation quality and integration quality in the user evaluation study.

%% file: Sections/6_UserEvaluation.tex
\section{User Evaluation}
We conducted a within-subject study with 12 BLV viewers to evaluate the effectiveness of DanmuA11y compared to a baseline. \hl{Specifically, we aim to answer the following research questions:}

\hl{(RQ1) \textbf{System Usability}: How do BLV viewers perceive the usability of DanmuA11y?}

\hl{(RQ2) \textbf{Danmu Comprehension}: How does DanmuA11y impact BLV viewers' Danmu comprehension?}

\hl{(RQ3) \textbf{Video Viewing Experience}: How effectively does DanmuA11y provide a smooth viewing experience?}

\hl{(RQ4) \textbf{Social Connection}: How does DanmuA11y affect BLV viewers' sense of co-watching with other viewers?}

\subsection{Participants and Materials}
\subsubsection{\textbf{Participants}}
We recruited 12 BLV viewers (P9-P20; seven male, five female, Table~\ref{tab:demographics} lists their demographics) who regularly watched videos on Danmu-enabled platforms. These participants were recruited from an online support community, with ages ranging from 22 to 41 (mean = 30.8, SD = 5.3). Eight participants were totally blind and four participants were legally blind with light perception. All participants had prior experience accessing Danmu on video platforms, with a viewing frequency ranging from two or three times a week to daily. \hl{None of the participants took part in the formative study.} Additionally, all participants were iOS VoiceOver users, had experience listening to spatial audio via headphones, and had normal hearing.

\subsubsection{\textbf{Apparatus}}
Each participant used two systems: DanmuA11y and a baseline. The baseline was the design probe used in the formative study, which simulated their current practices. Both systems were run on an iPhone 15. While using each system, participants wore AirPods Pro (second generation) with Personalized Spatial Audio\footnote{This feature personalizes the spatial audio to each participant's head-related transfer function (HRTF). https://support.apple.com/en-us/102596} enabled. Each participant adjusted VoiceOver's speech rate and audio ducking according to their preferences.

\subsubsection{\textbf{Videos}}
To ensure a fair comparison between the two systems, we selected six videos (V3-V8) from our 18-video dataset (Table~\ref{tab:video_evaluation}) and split them into two groups with comparable speech ratios, length, and content. Each group included three videos: an educational video (speech ratio > 95\%, length $\approx$ 2 minutes, featuring people explaining the scientific principles behind everyday tricks), a comedic video (speech ratio $\approx$ 80\%, length $\approx$ 5 minutes, featuring people humorously trying snacks), and a music video (speech ratio = 0\%, length $\approx$ 5 minutes, featuring instrumental music). Additionally, participants watched another two videos (V1-V2) during the tutorial phase and ten videos (V9-V18) after the comparison study.

\subsection{Design and Procedure}

\subsubsection{\textbf{Procedure}}
We first collected participants' demographics and asked about their prior experiences with Danmu. They then received a 10-minute tutorial on both systems, using the same two videos (V1-V2).

\begin{figure*}[h]
    \centering
    \includegraphics[width=\textwidth]{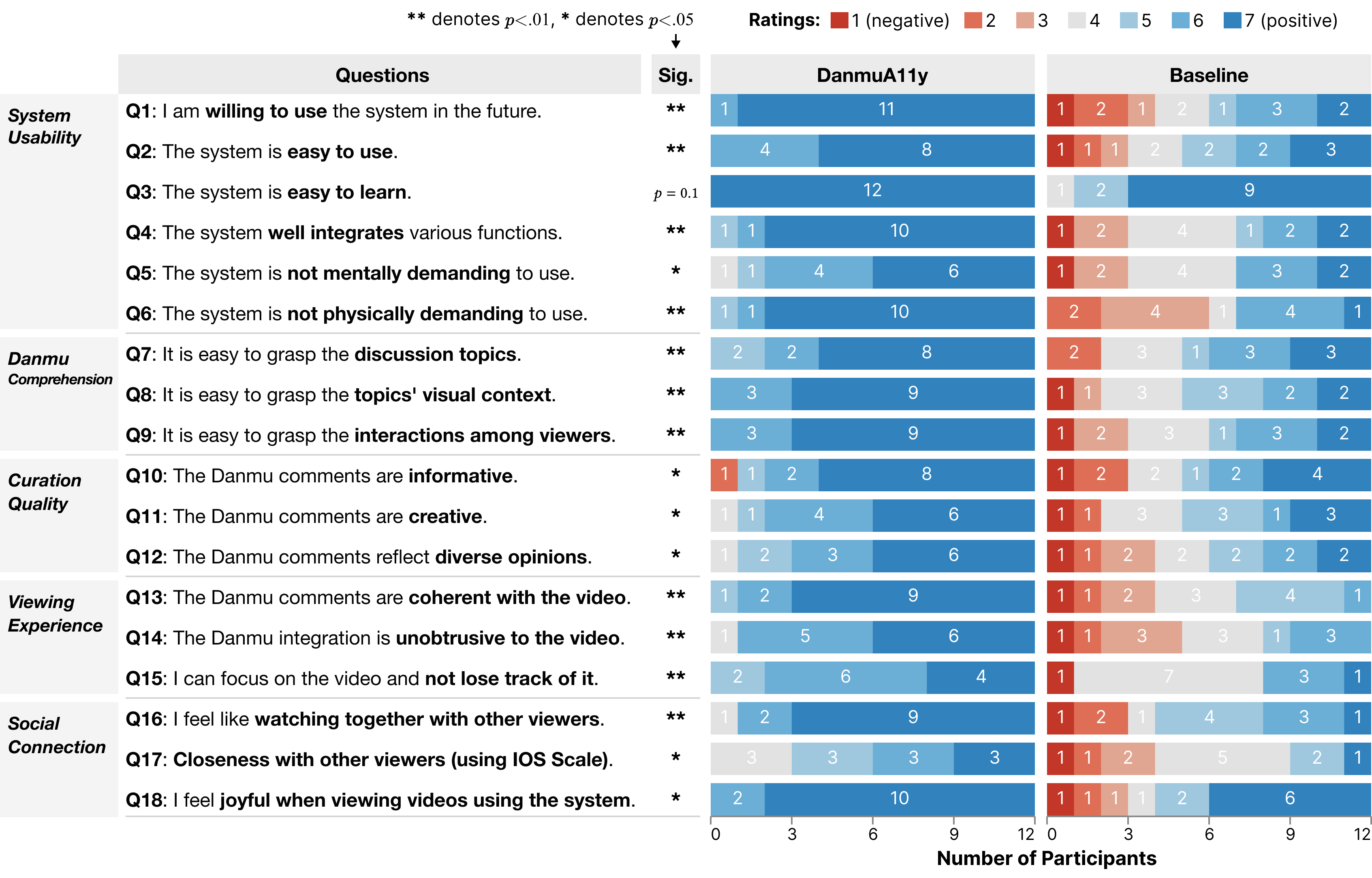}
    \caption{Distributions of the ratings for DanmuA11y and the baseline (1=strongly negative, 7=strongly positive). The asterisks indicate the statistical significance as a result of the Wilcoxon signed-rank test. Detailed statistics are provided in Table~\ref{tab:subjective_ratings}.}
    \label{fig:comparison_ratings}
    \Description{This figure shows the distributions of the ratings for DanmuA11y and the baseline. The figure shows that DanmuA11y outperformed the baseline in system usability, Danmu comprehension, curation quality, viewing experience, and social connection.}
\end{figure*}

After the tutorial, participants proceeded with the comparison study. Each participant used both systems to consume Danmu-embedded videos. To ensure study control, the order of systems and video groups was counterbalanced by creating four combinations, which were evenly assigned to the twelve participants. The videos were presented in random order. 
Participants were instructed to watch videos as they would in daily life (e.g., fast-forward, rewind, or pause at any time), to engage with Danmu as long as it did not disrupt their viewing experience, to \textbf{report any confusion} about Danmu, and to complete each video unless the system was frustrating. After watching each video, participants wrote a short \textbf{video summary}. Upon completing both systems, participants completed a questionnaire, with questions shown in Figure~\ref{fig:comparison_ratings}. A semi-structured interview was then conducted to gather participants' feedback.

After the comparison study, participants watched the remaining videos (V9-V18) using only DanmuA11y and were asked to think aloud about their viewing experiences. At the end of the study, participants were invited to share any suggestions or concerns they had regarding the system. The entire study lasted three hours and was conducted one-on-one, in-person. Participants were compensated approximately 34 USD in local currency for their time.

\subsubsection{\textbf{Metrics}}
We assessed both systems using comprehension metrics and subjective ratings.

\textbf{Comprehension Metrics.} 
To assess Danmu comprehension, we followed prior work \cite{wang2021toward}, tallying the \textit{instances of confusion} about Danmu reported by participants while viewing the video. For video comprehension, we adopted the method in \cite{van2024making}, analyzing the \textit{number of errors} present in participants' video summaries.

\textbf{Subjective Ratings.} We adapted established questionnaires \cite{lewis2018system,hart2006nasa,mankoff2003heuristic,aron1992inclusion} to investigate the four research questions. For usability (RQ1), we assessed participants' willingness to use the system in the future, ease of use, and ease of learning using relevant questions from the System Usability Scale \cite{lewis2018system}. 
For Danmu comprehension (RQ2), participants rated how easy it was to comprehend three key aspects of Danmu: the discussion topics, the visual context of those topics, and the interactions among viewers. 
To assess the video viewing experience (RQ3), we examined whether the Danmu-video integration caused minimal disruption, following the ambient display heuristics developed by Mankoff et al. \cite{mankoff2003heuristic}. For social connection (RQ4), we measured participants' sense of closeness with other viewers using the Inclusion of Other in the Self (IOS) Scale \cite{aron1992inclusion} and whether they felt as if they were co-watching videos with others \cite{huang2024sharing}. All questions used a 7-point Likert scale.

\subsubsection{\textbf{Analysis}}
We recorded the study's audio, tracked participants' interaction histories, and collected their questionnaire responses. We used the Wilcoxon signed-rank test \cite{wilcoxon1970critical} to analyze the significance of the questionnaire responses. We tallied instances where participants expressed confusion about Danmu during video viewing. To analyze the video summaries written by participants, one researcher first shuffled the summaries for each video to obscure participant and system information. Another researcher, who was unaware of these details, identified and labeled any factual errors in the summaries. The interview audio was transcribed and categorized according to the research questions. Our findings are reported based on this analysis.

\section{Evaluation Results}\label{sec:results}
During the comparison study, a total of 36 trials (12 participants $\times$ 3 videos) were conducted for each system. Participants completed all 36 trials using DanmuA11y. However, seven participants did not complete a total of nine trials\footnote{The uncompleted trials are listed in Table~\ref{tab:trials}.} when using the baseline due to frustration caused by the system. \hl{In the following, we report the results regarding the four RQs: system usability, Danmu comprehension, video viewing experience, and social connection.}

\subsection{System Usability (RQ1)}
\textbf{All participants preferred using DanmuA11y over the baseline for accessing Danmu while watching videos.} Participants reported a significantly higher willingness to use DanmuA11y in the future compared to the baseline ($Z=-2.82,\ p<.01$). They found DanmuA11y significantly easier to use ($Z=-2.70,\ p<.01$), with reduced mental effort ($Z=-2.54,\ p<.05$) and physical effort ($Z=-2.83,\ p<.01$). All participants rated DanmuA11y the highest score for ease of learning, because the system incorporated familiar design elements they were accustomed to, such as audio discussions mimicking ``\textit{group calls}'' (P11, P20) and notification sounds resembling ``\textit{new message alerts}'' (P9, P17). Additionally, participants rated DanmuA11y on how easy it was to distinguish between three sound sources—the original video, the AI narrator, and the human viewers—on a scale from 1 (very hard) to 7 (very easy). The average rating of 6.92 indicates that DanmuA11y was effective in helping participants differentiate between these sources using spatial audio and varied tones.

\textbf{DanmuA11y was particularly praised for  its minimal cognitive and physical demands.} Participants felt ``\textit{totally relaxed}'' (P16) when using DanmuA11y. They highlighted the Danmu insertion as ``\textit{seamless}'' (P12) and praised the on-demand access to Danmu via the shake gesture as ``\textit{effortless}'' (P19), ``\textit{convenient}'' (P14), and ``\textit{flexible}'' (P15), as it eliminated the need for ``\textit{multi-step, repetitive operations required by screen readers}'' (P13). In contrast, the baseline made participants feel like they were ``\textit{multi-tasking}'' (P20), with P12 explaining, ``\textit{I had to constantly tap the screen, digest disorganized information, manage concurrent speeches, and shift my focus between the video and comments. It took the enjoyment out of watching videos.}''

\subsection{Danmu Comprehension (RQ2)}
In terms of Danmu comprehension, DanmuA11y outperformed the baseline in both comprehension metrics and subjective ratings. The results are as follows.

\subsubsection{\underline{\textbf{Comprehension Metrics}}}
\textbf{DanmuA11y significantly reduced confusion about Danmu reported by participants} compared to the baseline ($Z=-2.94,\ p<.01$). With the baseline, the twelve participants reported a total of 52 instances of confusion (1.44 times per video). These instances were caused by missing visual context (31 instances),  unfamiliar Internet slang (6 instances,  e.g., the name of video games), near-homophones (3 instances), and incomplete dialogues (12 instances). For example, P11 expressed confusion over the comment, ``\textit{The one mentioning `meat', are you a devil?}'' because preceding comments that provided necessary context were absent. In contrast, when using DanmuA11y, participants only reported nine instances of confusion (0.25 times per video) . These cases were caused by near-homophones (5 instances, e.g., P10 misheard ``\textit{hospital, `yee-yuan'} '' as its Chinese near-homophone ``\textit{music, `yin-yweh'} '') and unfamiliar Internet slang (4 instances). These results demonstrate that DanmuA11y effectively reduced confusion by providing visual descriptions and organizing comments into dialogues among viewers.

\begin{figure*}[h]
    \centering
    \includegraphics[width=\textwidth]{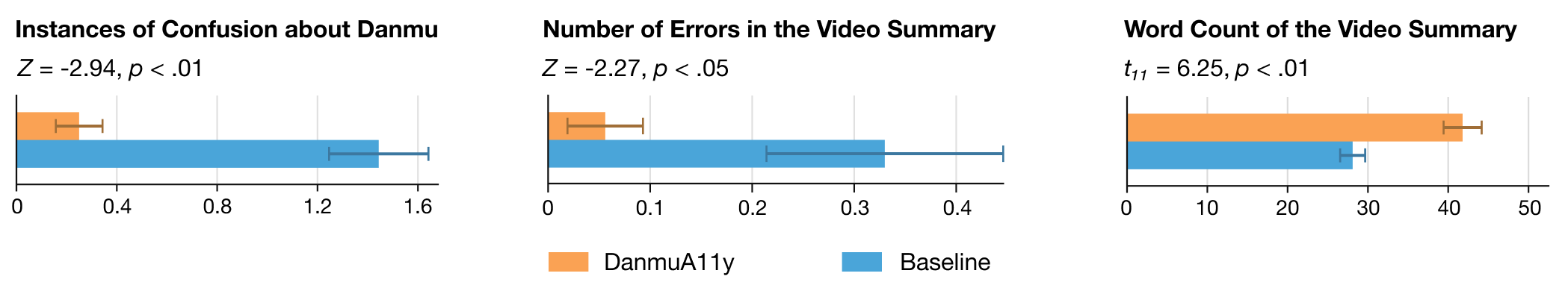}
    \caption{Comprehension metrics in the evaluation study. These metrics were calculated as \textbf{averages for each participant per video}. Error bars indicate standard errors.}
    \label{fig:comprehension_metrics}
    \Description{This figure shows the comprehension metrics in the evaluation study. It shows that DanmuA11y outperformed the baseline with reduced confusion about Danmu, reduced errors in video summaries, and increased word count of the video summaries.}
\end{figure*}

\textbf{DanmuA11y improved participants' video comprehension.} After using DanmuA11y, participant-written video summaries contained significantly more words than the baseline ($t_{11}=6.25,\ p<.01$)\footnote{We used a paired t-test for significance analysis, as the data satisfied the assumptions of normality and homogeneity of variances according to the Shapiro-Wilk Test and Levene's Test, respectively.}. On average, participants wrote 41.8 words per video with DanmuA11y, compared to 28.1 words with the baseline. 
Moreover, participants made significantly fewer factual errors in the video summaries with DanmuA11y ($Z=-2.27,\ p<.05$). When using the baseline, they made a total of 14 factual errors (0.33 times per video), of which 11 were visual errors resulting from ``\textit{guessing visuals using comments as clues}'' (P9). In contrast, with DanmuA11y, the total number of errors were reduced to just two instances (0.06 times per video). Participants attributed their improved video comprehension to the inclusion of visual descriptions (mentioned 11 times) and their ability to focus more attentively on the videos (seven times).

\subsubsection{\underline{\textbf{Subjective Ratings}}}

\textbf{Participants reported that DanmuA11y significantly enhanced Danmu comprehension} across three key aspects: discussion topics ($Z=-2.69,\ p<.01$), visual context ($Z=-2.83,\ p<.01$), and interactions among viewers ($Z=-2.70,\ p<.01$). Participants noted that organizing comments into dialogues ``\textit{made the comments much easier to understand}'' (P13) and helped them ``\textit{follow the conversations}'' (P12).

\textbf{Participants highlighted that visual descriptions were essential for deeply engaging with Danmu.} For example, after watching V8, P10 remarked, ``\textit{Without the description, `a cat lying on the musical instrument', I would've been quite confused. Why did people mention cats in a music video? Thanks to that description, I didn't feel excluded from the discussion.}''  Participants appreciated how AI-generated descriptions complemented viewer comments, offering a more comprehensive understanding. As P12 observed, ``\textit{When others said `It's like eating concrete', the AI description promptly addressed my curiosity: `a plate of brown bricks'. These pieces came together into a full picture.}'' Additionally, participants noticed errors in visual descriptions when they conflicted with the video's audio, such as when the description ``\textit{A man dancing}'' contradicted the audio ``\textit{When you walk outdoors}'' in V6. However, they did not notice errors when there was no such conflict, which aligns with issues reported in previous research \cite{huh2023avscript,van2024making}. To address this issue, future work should focus on enhancing visual description accuracy or providing descriptions from multiple sources \cite{lee2021image}.

\textbf{DanmuA11y presented comments of significantly higher quality compared to the baseline.} Participants found the comments in DanmuA11y to be significantly more informative ($Z=-2.21,\ p<.05$), creative ($Z=-2.56,\ p<.05$), and reflective of diverse opinions ($Z=-2.51,\ p<.05$). For example, P10 found the comment ``\textit{(the candy bar is) like my dog's chew toy}'' to be both creative and helpful in visualizing the snack. Similarly, P11 noted, ``\textit{It's so funny when someone says, `Coach, I want to learn', and another person replies, `Coach: Why don't you learn everything?' It (DanmuA11y) really helped me discover these creative interactions.}'' In contrast, participants described the comments in the baseline as ``\textit{chaotic}'' (P15) and ``\textit{messy}'' (P17), which made it challenging to ``\textit{get key points and find highlights}'' (P18). These results indicate that DanmuA11y effectively curated content according to the three criteria. 
Additionally, participants suggested incorporating more curation criteria to better personalize the content based on their preferences, which we discuss in Section~\ref{sec:personalization}.

\subsection{Video Viewing Experience (RQ3)}

\textbf{DanmuA11y offers a smooth video viewing experience.} Participants rated the Danmu comments in DanmuA11y as significantly more coherent with the video ($Z=-3.08,\ p<.01$) and unobtrusive ($Z=-2.82,\ p<.01$) compared to the baseline. They described the viewing experience as ``\textit{smooth}'' (P16) and noted that ``\textit{the seamless blend of Danmu and the videos creates a cohesive piece of work}'' (P10). Participants also appreciated that the notification sound during speech breaks was ``\textit{unobtrusive}'' (P11) and ``\textit{natural}'' (P9), as ``\textit{the sound didn't disrupt the speech flow}'' (P13, P18).

\textbf{The Danmu-video integration in DanmuA11y resulted in minimal split attention.} Participants reported that DanmuA11y allowed them to stay more focused on the video and not lose track of it ($Z=-2.75,\ p<.01$), thanks to the absence of overlapping speech (twelve times), the well-timed comments (nine times), and the auto-rewinding feature (ten times, described in Section~\ref{sec:on-demand-design}). As P10 explained, ``\textit{The video automatically rewound to the point where the bubble sound occurred, not where I shook the phone. This is ingenious because it allowed me to refocus on the video effortlessly, without the trouble of manually adjusting the progress bar.}''

\hl{\textbf{DanmuA11y enabled participants to easily access high-quality comments while watching videos.} On average, participants accessed more comments using DanmuA11y (51 comments per video) compared to the baseline (33 comments per video). This improvement was attributed to DanmuA11y's simple interactions (P12) and high-quality content (P17), which fostered user engagement with comments. In contrast, the baseline required repetitive and tiring operations, preventing participants from accessing comments. For example, participants had to repeatedly tap the screen every few seconds to ``\textit{keep up with the comment list}'' (P14). These repetitive actions caused fatigue within one to two minutes, leading participants to either stop reading the comments or end the trial early. 
P12 succinctly expressed this frustration: ``\textit{The frequent screen reader operations made me tired. Plus, the comments were so chaotic that I couldn’t enjoy the video.}'' These findings highlight the importance of enabling easy access to high-quality comments to provide an enjoyable viewing experience.}

\hl{\textbf{Participants exhibited varied usage patterns with DanmuA11y, indicating directions for personalization.} DanmuA11y offers user control through several features: on-demand Danmu access via shake gestures, an Audio-Discussion toggle, and video playback controls. We observed personalized usage patterns for these features:}

\hl{\textbf{(1) On-demand Danmu Access}: On average, participants responded to 87.4\% (SD=9.2\%) of Danmu notifications by performing shake gestures, indicating that they accessed most on-demand comments. Participants typically ignored new comments when previous ones did not match their preferences. For instance, P15 skipped the last two notifications in V3, explaining, ``\textit{I found previous comments in this video less informative than I wanted, so I skipped the last ones.}'' This finding highlights the opportunity to refine comment curation according to individual preferences.}

\hl{\textbf{(2) Audio-Discussion Toggle}: Participants primarily used the audio-discussion toggle to control their engagement with audience discussions. For example, nine participants disabled the toggle during the chorus of an instrumental music video (V7) to ``\textit{focus on the best part of the music}'' (P9) and re-enabled it afterward to ``\textit{see what other people think of it}'' (P14). This behavior suggests the potential for tailoring Danmu presentation based on video types—such as minimizing comments during key moments in music videos or increasing discussions for controversial news videos.}

\hl{\textbf{(3) Playback Control}: Participants used playback controls according to their daily habits. Eight participants did not manually rewind videos, while the other four participants rewound an average of 1.42 times per video. When participants rewound, their goals included ``\textit{revisiting funny comments}'' (P16) and ``\textit{double-checking video details such as food names}'' (P20). These varied usage patterns highlight the importance of providing flexible controls to accommodate individual needs, which we further discuss in Section~\ref{sec:personalization}.}

\subsection{Social Connection (RQ4)}
In terms of social connection, DanmuA11y significantly enhanced participants' sense of social presence, connection, and enjoyment compared to the baseline. The results are as follows.

\textbf{DanmuA11y significantly enhanced the sense of watching together with other viewers} as indicated by higher ratings of social presence compared to the baseline ($Z = -2.83, \ p < .01$). As P11 noted, ``\textit{It really felt like there’s a group of people watching with me.}'' Participants attributed this co-watching experience to several features of DanmuA11y, particularly spatial audio and alternating tones. P18 remarked, ``\textit{The alternating tones on my left and right immediately gave me the impression of several people conversing nearby.}'' This finding aligns with prior research \cite{immohr2023proof}, which indicates that spatial audio improves social presence in remote communication. Moreover, the dialogue-like organization of the comments further contributed to the shared viewing experience. P11 described the comment organization as ``\textit{a group of people sharing their funny thoughts about videos}'',  which made the experience ``\textit{truly feel like co-watching with friends}'' (P14).

\textbf{This co-watching experience also increased participants' sense of connection with other viewers}, with higher ratings of closeness compared to the baseline ($Z = -2.55, \ p < .05$). Participants reported feeling deeply connected with other viewers, especially when the comments reflected their own thoughts or interests. For instance, P12 noted, ``\textit{It’s a pleasant surprise when someone said exactly what I was thinking.}'' Similarly, P18 mentioned, ``\textit{I'm really happy when I hear others mention my favorite movies—It’s like we're emotionally in sync.}'' Such emotional resonance not only made participants feel as if they were ``\textit{meeting someone who understands me well}'' (P11), but also ``\textit{added extra joy to the viewing experience}'' (P17).

\textbf{Consequently, participants reported greater enjoyment when using DanmuA11y} compared to the baseline ($Z = -2.23, \ p < .05$). In addition to ``\textit{the emotional resonance with other viewers}'' (P11), participants highlighted sources of joy, such as ``\textit{the creative comments}'' (P9) and ``\textit{the co-watching atmosphere}'' (P16: ``\textit{Watching with others makes me happy. I'm no longer alone; I'm accompanied.}''). Moreover, DanmuA11y enhanced user enjoyment by ``\textit{filling the awkward silence}'' (P16). As P16 explained, ``\textit{Previously, when I didn't get a joke in videos, it felt boring and awkward. But now, other people actually fill that awkward silence and make me laugh.}'' Overall, the results indicate that by transforming Danmu into multi-viewer discussions, DanmuA11y effectively created an enjoyable and socially engaging experience for BLV video viewers.

%% file: Sections/7_Discussion.tex
\section{Discussion}
Our formative study revealed three key challenges that hinder BLV viewers' engagement with Danmu: the lack of visual context, the speech interference between comments and videos, and the disorganization of comments. 
To address these challenges, we designed DanmuA11y with three core features: augmenting Danmu with visual context, seamlessly integrating Danmu into videos, and presenting Danmu via multi-viewer discussions. 
User evaluations demonstrated that DanmuA11y effectively improved Danmu comprehension, offered smooth video viewing experiences, and fostered social connections for BLV viewers. In the following sections, we discuss: (1) directions for personalizing DanmuA11y, (2) generalizability of DanmuA11y across videos, (3) integration of DanmuA11y with existing accessibility techniques, and (4) implications for enhancing commentary accessibility in broader contexts.

\subsection{Personalization of DanumA11y}\label{sec:personalization}
Based on the evaluation study, we identify several directions for tailoring DanmuA11y to different user contexts:

\textbf{Personalize the weights of curation criteria.} While DanmuA11y assigns equal importance to its three curation criteria, participants expressed individualized preferences. For example, P15 favored informative comments, while P16 preferred humorous or creative comments. To better align with personal preferences, future systems could allow users to adjust the weight of each criterion, or adapt to user preferences by learning from users' interaction histories \cite{hoi2021online}.

\textbf{Tailor curation to different video types.} Participants' preferences for comments also varied depending on the video type. For instance, humorous comments were favored for comedic content (V3-V4), diverse opinions were valued for news videos (V9-V10), and informative comments were preferred for action movies with complex visuals (V11-V12). This suggests the potential to fine-tune comment curation based on the video types.

\textbf{Incorporate user-defined content filters.} Participants proposed adding more filtering options, such as avoiding spoilers (P20), highlighting replies to the video creators' questions (P14), prioritizing comments posted by creators (P19), and emphasizing comments with many likes (P12). Future systems could better address diverse user needs by supporting user-defined filters \cite{jhaver2022designing} or question-answering interactions \cite{antol2015vqa,kwiatkowski2019natural}.

\textbf{Enable flexible control in daily scenarios.} Currently, DanmuA11y relies on phone-shake gestures for on-demand access. To improve usability in everyday situations, future systems could support alternative input methods, such as head gestures using earphones \cite{yan2018headgesture,yan2020headcross}, or hand gestures with hand-worn devices \cite{xu2022enabling,gu2019accurate}.

\textbf{Support acoustic feature customization.} Participants suggested customizable acoustic features, such as adjusting the speech rate, tone, comment density, and adding reverberation to simulate different environments (e.g., a theater). Additionally, future work could explore more diverse auditory representations \cite{guerreiro2023design,may2020spotlights}, such as utilizing dynamic sound sources to simulate the presence of moving viewers.

\subsection{System Generalizability Across Videos}\label{sec:generalizability}

\hl{DanmuA11y was evaluated using 18 videos with diverse visual styles and speech ratios. Based on the evaluation results, we reflect on the generalizability of DanmuA11y and identify potential directions for improvement.}

\subsubsection{\hl{\textbf{AI Visual Description.}}} 
The 18 videos covered a wide range of visual styles, from nearly static imagery (e.g., V18) to rapidly changing scenes (e.g., V11). When generating visual descriptions, the pipeline achieved an accuracy of 92.5\%, suggesting that the AI descriptions were accurate for most cases. However, there are several areas for improvement. 

\textbf{The first area for improvement is describing complex visual changes}, such as the fight scenes in V11 and V12. Because the current pipeline uses key frames for generating descriptions, it cannot capture movement over time and can lead to incorrect descriptions of human actions. To address these issues, future work could incorporate recent advances in video-to-language models \cite{cheng2024videollama,jin2024chat}, utilize vision-language pre-training methods \cite{wang2023large,yang2019cross}, or explore specialized pipelines, such as reconstructing 3D meshes of rapidly changing objects and describing their movements \cite{delmas2022posescript}. 

\textbf{The second area for improvement is filtering out similar descriptions across topics}. Our pipeline generates different descriptions when Danmu topics discuss different visual elements (e.g., hair, clothing, or facial expressions in V18). However, when the topics center on the same visual element, the descriptions become similar. Future systems could address this by removing similar descriptions through methods like semantic comparison \cite{reimers2019sentence}. 

\textbf{Additionally, the coverage of AI descriptions could be enhanced.} The current AI descriptions aim to provide visual context for comments but may not capture the most salient visual objects. They are also not as comprehensive as audio descriptions. To enhance the coverage of AI descriptions, future work could use saliency detection methods \cite{liu2021visual} or integrate DanmuA11y with existing audio descriptions, which we discuss in Section~\ref{sec:integration}.

\subsubsection{\hl{\textbf{Danmu-Video Integration.}}} 
\hl{The current pipeline restricts Danmu placement to non-speech segments or speech breaks. For videos with few speech segments (e.g., V9, with 100\% speech), most comments were added at speech breaks, which could result in frequent notifications. To address this, future work could explore extending the video timeline with AI-generated background music \cite{copet2024simple} to create more spaces for Danmu insertion.}

\subsubsection{\hl{\textbf{Audio Presentation.}}}
\hl{DanmuA11y utilizes spatial audio and varied human tones to present Danmu as multi-viewer discussions. However, some videos may have used similar spatial-audio layouts or speech tones, which may lead to confusion. Future work could explore detecting similar audio \cite{li2017comparison} and adaptively adjusting the audio presentation (e.g., using other tones) to ensure clear differentiation between Danmu and videos.}

\subsection{Integration with Existing Techniques}\label{sec:integration}

\hhl{Based on observations and user feedback from the evaluation study, we discuss how DanmuA11y could be further integrated with existing accessibility techniques.}

\subsubsection{\hl{\textbf{Audio Descriptions.}}}
\hhl{Participants in our study valued the AI visual descriptions for providing additional visual context. However, some participants expressed a preference for professional audio descriptions (AD) due to their higher accuracy and better alignment with the video content. To address this, DanmuA11y could be enhanced by integrating ADs as part of the video input. Based on user feedback and video accessibility guidelines \cite{pavel2020rescribe}, we suggest three considerations for this integration: (1) \textit{Avoiding redundant descriptions}: the AI visual descriptions should provide provide new, complementary information rather than repeating the ADs. (2) \textit{Minimizing speech interference}: Comments should not overlap with speech from the video or the ADs. This can be achieved by placing comments at speech breaks or by extending the video timeline with looping background music, a method accepted by BLV viewers \cite{pavel2020rescribe}. (3) \textit{Distinguishing different content}: To help users differentiate between the video, ADs, and Danmu comments, these elements could be optionally presented using distinct spatial positions or varying audio tones.}

\subsubsection{\hl{\textbf{Screen Readers.}}}
\hhl{In our formative study, participants reported that using screen readers to access the fast-scrolling list of comments was tedious, as it required repetitive touch input to retrieve the latest comments. To address this, DanmuA11y introduced a shaking feature, enabling users to access curated comments more easily. User evaluations showed that all participants found this feature more effortless and convenient compared to screen reader access. However, one participant (P17) noted that the feature lacked the flexibility to navigate between comments, which is particularly useful for ``\textit{skipping or revisiting comments}''. To provide users with flexible access to the curated comment list, future systems could consider also presenting the list through screen readers. This approach would allow viewers to easily access on-demand comments by shaking their phones while maintaining the agency to navigate between curated comments via screen reader controls.}

\subsection{Broader Implications for Commentary Accessibility}
By addressing Danmu accessibility, our research offers insights for improving accessibility in broader contexts, especially in scenarios involving real-time or time-synced comments.

For \textbf{real-time commentary in live-streaming} \cite{jun2021exploring,li2022feels}, prior research indicates that BLV users face challenges in keeping up with rapidly updating comments and understanding the discussions among other viewers \cite{jun2021exploring}. Our design insights suggest potential directions for enhancing commentary accessibility in live-streaming: (1) supplementing comments with visual context, such as descriptions of relevant visual objects in the stream; (2) curating comments based on users' information preferences; (3) avoiding overlaps with the streamer's speech, for instance, by pausing comments when the streamer is speaking; and (4) using spatial audio and varied tones to create a co-watching experience.

\hl{To implement the proposed design, future work needs to address both \textbf{technical and human-factor challenges}. The key technical challenge lies in achieving real-time processing, which involves several aspects: generating visual descriptions (e.g., via real-time computer vision models \cite{chang2024worldscribe,cheng2024yolo}), curating comments (e.g., with pre-trained sentence encoders \cite{reimers2019sentence,cer2018universal}), identifying non-speech periods (e.g., using human speech detection algorithms \cite{ramirez2004efficient,tanyer2000voice}), and optimizing comment placements (e.g., sorting comments in descending order of quality). 
Regarding the human-factor challenge, live-streaming often involves two-way communications between streamers and their audiences~\cite{lu2018you}. This interaction might create unique information needs, such as prioritizing comments that respond to live-streamers' questions, filtering out abusive remarks \cite{roy2023hateful}, or removing promotional messages that disrupt the stream's flow \cite{chen2021afraid}. Further research is necessary to identify these specific information needs.} 

In \textbf{video-based social media} platforms without Danmu, our multi-viewer discussion design could be adapted by using time-referenced comments \cite{yarmand2019can} (e.g., ``\textit{2:50 this kid is soooo funny}'') or by aligning traditional comments with the video timeline \cite{wehrmann2020adaptive} to create a co-watching experience for BLV viewers. \hl{Moreover, DanmuA11y's key design insight—facilitating easy access to high-quality content—can be applied to broader contexts like audio descriptions. To balance limited video time with the need for detailed descriptions \cite{van2024making,ning2024spica}, future audio descriptions could insert essential visual information into non-speech segments and provide supplementary explanations during speech breaks. Users could access these additional details through simple interactions like shake gestures, allowing for a seamless viewing experience while retaining the option for more content.}

\subsection{Limitations and Future Work}
DanmuA11y is designed to make Danmu consumption accessible to BLV viewers. Future research should explore methods to support BLV users in composing and sharing their own Danmu comments, which may lead to new needs for Danmu consumption, such as curating comments that resemble their own posts. Our work focuses on improving video viewing experiences via auditory feedback. Future works could explore multi-sensory stimuli—such as vibrations \cite{tactileCompass,virtualpaving} or lights \cite{lightguide}—to further enhance user engagement. 
While our work focuses on using Danmu for entertainment and social interaction, future studies could explore the accessibility of Danmu for other applications, such as enhancing online learning outcomes \cite{chi15learningDanmu} or assisting BLV content creators in gathering viewer feedback \cite{chen2022danmuvis}. In addition to using automatic approaches, future work could leverage social support to further enhance Danmu accessibility. For instance, providing automatic description suggestions to sighted viewers could encourage them to include visual descriptions in their comments, thereby benefiting BLV users.

The current pipeline is designed to process existing comments. To keep the system updated with new Danmu comments, it could classify new comments into existing topics \cite{pham2023topicgpt} or periodically re-run the pipeline. The optimization algorithm does not account for interactions between topics, such as re-ordering them to form a coherent sequence, which could be considered in future work. Our use of GPT-4o for topic modeling \cite{pham2023topicgpt}, image captioning \cite{hu2023promptcap}, and creativity rating \cite{park2023generative} occasionally results in hallucinations and errors. Future work could address these issues by incorporating specialized models \cite{vayansky2020review,campos2023machine,sun2023new} or model advancements. Beyond generating visual descriptions to enhance Danmu accessibility, future research could explore how to leverage Danmu comments to improve video accessibility \cite{wang2021toward}, such as using comments to generate creative audio descriptions \cite{walczak2017creative}. Additionally, exploring the long-term impact of DanmuA11y on the social engagement of BLV viewers presents a promising direction for future research.

%% file: Sections/8_Conclusion.tex
\section{Conclusion}
DanmuA11y is a system designed to improve the accessibility of Danmu for BLV viewers. 
It addresses three primary challenges: the lack of visual context, the speech interference between comments and videos, and the disorganization of comments. Through user evaluations, we demonstrated that DanmuA11y effectively improved Danmu comprehension, provided a smooth viewing experience, and fostered social connections among viewers. We also identified future directions for personalizing DanmuA11y to accommodate diverse user needs and distilled implications for improving commentary accessibility in video and live-streaming platforms. We hope this work offers valuable insights for enhancing social media accessibility and inspires researchers to develop tools that facilitate social engagement for BLV users.

%% file: Tables/demographics.tex
\begin{table*}[!htbp]
  \centering
  \caption{Participants' demographics. P1-P8 participated in the formative study, and P9-P20 participated the evaluation study.}
  \renewcommand{\arraystretch}{1.3}

  \setlength{\tabcolsep}{1mm}{
      \begin{tabular}{c c c l l l l l}
        \toprule
        \textbf{PID} & \textbf{Age} & \textbf{Gender} & \textbf{Visual Condition} & \textbf{Danmu-enabled Video Platforms} & \textbf{Experience} & \textbf{Usage Frequency} \\
        \midrule
        P1 & 27 & F & Legally blind & Bibibili, Douyin, Youku & 4 years & Daily \\
        P2 & 34 & M & Totally blind & Bibibili, Youku & 1 year & Daily \\
        P3 & 26 & M & Legally blind & Bibibili, Douyin & 3 years & 2-3 times per week \\
        P4 & 27 & M & Totally blind & Douyin, Youku & 3 years & 4-5 times per week \\
        P5 & 38 & F & Totally blind & Bibibili, Douyin  & <1 year & Daily \\
        P6 & 35 & F & Legally blind & Bibibili & 4 years & 1 time per week \\
        P7 & 30 & F & Legally blind & Bibibili, Tencent & 1 year & 2-3 times per week \\
        P8 & 33 & M & Totally blind & Douyin & 3 years & Daily \\
        P9 & 22 & M & Totally blind & Douyin, Youku & 2 years & Daily \\
        P10 & 30 & F & Totally blind & Bibibili, Douyin & 1 year & 2-3 times per week \\
        P11 & 32 & F & Totally blind & Bibibili, Douyin, iQiyi, Tencent & 3 years & Daily \\
        P12 & 41 & M & Legally blind & Tencent, Douyin & 4 years & Daily \\
        P13 & 32 & M & Totally blind & Bibibili, Douyin, Youku & 5 years & 4-5 times per week \\
        P14 & 36 & M & Totally blind & Bibibili, Youku & 2 years & 2-3 times per week \\
        P15 & 28 & F & Legally blind & Bibibili, Douyin & 4 years & 2-3 times per week \\
        P16 & 23 & M & Totally blind & Bibibili, Douyin, iQiyi & 1 year & Daily \\
        P17 & 36 & M & Totally blind & Tencent, iQiyi & 4 years & Daily \\
        P18 & 30 & F & Legally blind & Bibibili, Douyin & 2 years & 4-5 times per week \\
        P19 & 31 & M & Totally blind & Bibibili, Douyin, iQiyi & 2 years & Daily \\
        P20 & 29 & F & Legally blind & Bibibili, Youku, iQiyi, Tencent & 3 years & Daily \\
        \bottomrule
      \end{tabular}
    }
  \label{tab:demographics}
\end{table*}

%% file: Tables/video_formative.tex
\begin{table}[htbp]
\centering
\caption{The six videos used in the formative study.}
\renewcommand{\arraystretch}{1.3}
\begin{tabular}{llc}
\toprule
\textbf{Type} & \textbf{Video Content} & \textbf{Length} \\
\midrule
{Educational} & {A person explains how fast humans can really run.} & {01:04} \\
{Comedic} & {A collection of funny bloopers.} & {03:17} \\
{Tutorial} & {A person shows how to prepare ``egg dishes'' for a meal.} & {04:28} \\
{News} & {A news clip of a three-year-old girl caught by a kite.} & {00:33} \\
{Music} & {A person sings the music ``Little Love Song''.} & {04:33} \\
{Film Clip} & {Dialogues from the dinner party scene in ``Eat Drink Man Woman''.} & {06:56} \\
\bottomrule
\end{tabular}
\label{tab:video_formative}
\end{table}

%% file: Tables/video_evaluation.tex
\begin{table}[htbp]
\centering
\caption{The 18-video dataset used in the evaluation study. \textbf{V3-V8 are used for the controlled comparison}.}
\renewcommand{\arraystretch}{1.3}

\begin{tabular}{lllcr}
\toprule
\textbf{ID} & \textbf{Type} & \textbf{Video Content} & \textbf{Length} & \textbf{Speech Ratio} \\
\midrule
V1  & Film Clip        & Dialogues between two people in a grocery store. & 2:04  & 48\% \\
V2  & News             & A sports commentary on a diving event.                          & 1:22  & 80\% \\
\textbf{V3}  & \textbf{Comedic}          & \textbf{Two people humorously try trending internet snacks.}              & \textbf{5:08}   & \textbf{83\%} \\
\textbf{V4}  & \textbf{Comedic}          & \textbf{One person humorously tries celebrity-recommended food.}     & \textbf{4:47}   & \textbf{83\%} \\
\textbf{V5}  & \textbf{Educational}      & \textbf{A person explains the proper way to cook dumplings.}              & \textbf{1:45}   & \textbf{97\%} \\
\textbf{V6}  & \textbf{Educational}      & \textbf{A person explains why shoes get wet on rainy days.}               & \textbf{1:48}   & \textbf{100\%} \\
\textbf{V7}  & \textbf{Music}            & \textbf{Instrumental music ``Battle for Prince of Lan Ling''.}             & \textbf{4:59}   & \textbf{0\%}  \\
\textbf{V8}  & \textbf{Music}            & \textbf{Instrumental music ``Senbonzakura''.}                         & \textbf{4:56}   & \textbf{0\%}  \\
V9 & News             & A report on the controversy surrounding self-driving cars.       & 2:48   & 100\% \\
V10 & News             & A report on an event during the Paris Olympics.        & 3:43   & 100\% \\
V11 & Film Clip        & A group of people fight using traditional Chinese kung fu.       & 4:19   & 17\% \\
V12 & Film Clip        & A group of people fight with each other using cameras.           & 5:33   & 9\%  \\
V13  & Tutorial         & A person demonstrates how to make a mosquito killer device.         & 5:58   & 66\% \\
V14 & Tutorial         & A person shows how to make a large ice-cream bar.             & 6:49   & 86\% \\
V15 & Tutorial         & A person demonstrates how to make pudding. & 2:47   &96\%  \\
V16 & Music            & Two people sing the music ``The Old Man and the Sea''.      & 4:23   & 64\% \\
V17 & Comedic          & A person tries to escape home with a funny excuse.   & 0:24   & 12\% \\
V18 & Educational      & The evolution of Chinese women’s hairstyles over the last century. & 3:55 & 12\% \\

\bottomrule
\end{tabular}
\label{tab:video_evaluation}
\end{table}

%% file: Tables/evaluation_stats.tex
\begin{table*}[!hbtp]
\caption{\centering Detailed statistics of subjective ratings from the evaluation study (1 = strongly negative, 7 = strongly positive). \newline SD denotes the standard deviation. Significance was analyzed using the Wilcoxon signed-rank test.}
\label{tab:subjective_ratings}
\centering
\resizebox{1.0\columnwidth}{!}{
\setlength{\tabcolsep}{1.32mm}{
\renewcommand\arraystretch{1.1}
\newcommand{\tabincell}[2]{\begin{tabular}{@{}#1@{}}#2\end{tabular}}
\newcommand{\hlineblack}{\specialrule{0.1em}{0em}{0em}}
\begin{tabular}{c | l | c | c | c | c | c}
\hlineblack
\multirow{2}{*}{\textbf{Aspects}} & \multirow{2}{*}{\textbf{Questions}} & \multicolumn{2}{c|}{\textbf{DanmuA11y}} & \multicolumn{2}{c|}{\textbf{Baseline}} & \multirow{2}{*}{\textbf{Significance}}\\
\cline{3-4} \cline{5-6}
& & \textbf{Mean} & \textbf{SD} & \textbf{Mean} & \textbf{SD} & \\

\hlineblack
\multirow{6}{*}{\parbox[c]{2.0cm}{\centering \hspace{2.5mm}System\newline Usability}} 
& Q1: I am \textbf{willing to use} the system in the future. & 6.92 & 0.29 & 4.42 & 2.07 & $Z=-2.82,\ p<.01$ \\
& Q2: The system is \textbf{easy to use}. & 6.67 & 0.49 & 4.75 & 2.01 & $Z=-2.70,\ p<.01$ \\
& Q3: The system is \textbf{easy to learn}. & 7.00 & 0.00 & 6.42 & 1.08 & $Z=-1.63,\ p=0.1$ \\
& Q4: The system \textbf{well integrates} various functions. & 6.75 & 0.62 & 4.50 & 1.78 & $Z=-2.82,\ p<.01$ \\
& Q5: The system is \textbf{not mentally demanding} to use. & 6.25 & 0.97 & 4.58 & 1.83 & $Z=-2.54,\ p<.05$ \\
& Q6: The system is \textbf{not physically demanding} to use. & 6.75 & 0.62 & 4.25 & 1.82 & $Z=-2.83,\ p<.01$ \\
\hline
\multirow{3}{*}{\parbox{2.0cm}{\centering \hspace{2.5mm}Danmu\newline Comprehension}} 
& Q7: It is easy to grasp the \textbf{discussion topics}. & 6.50 & 0.80 & 5.00 & 1.81 & $Z=-2.69,\ p<.01$ \\
& Q8: It is easy to grasp the \textbf{topics' visual context}. & 6.75 & 0.45 & 4.75 & 1.71 & $Z=-2.83,\ p<.01$ \\
& Q9: It is easy to grasp the \textbf{interactions among viewers}. & 6.75 & 0.45 & 4.67 & 1.83 & $Z=-2.70,\ p<.01$ \\
\hline
\multirow{3}{*}{\parbox{2.0cm}{\centering \hspace{2.5mm}Curation\newline Quality}} 
& Q10: The Danmu comments are \textbf{informative}. & 6.25 & 1.48 & 4.83 & 2.21 & $Z=-2.21,\ p<.05$ \\
& Q11: The Danmu comments are \textbf{creative}. & 6.25 & 0.97 & 4.75 & 1.91 & $Z=-2.56,\ p<.05$ \\
& Q12: The Danmu comments reflect \textbf{diverse opinions}. & 6.17 & 1.03 & 4.42 & 1.93 & $Z=-2.51,\ p<.05$ \\
\hline
\multirow{3}{*}{\parbox{2.0cm}{\centering Video Viewing\newline Experience}} 
& Q13: The Danmu comments are \textbf{coherent with the video}. & 6.67 & 0.65 & 3.92 & 1.44 & $Z=-3.08,\ p<.01$ \\
& Q14: The Danmu integration is \textbf{unobtrusive to the video}. & 6.33 & 0.89 & 3.92 & 1.62 & $Z=-2.82,\ p<.01$ \\
& Q15: I can focus on the video and \textbf{not lose track of it}. & 6.17 & 0.72 & 4.50 & 1.57 & $Z=-2.75,\ p<.01$ \\
\hline
\multirow{3}{*}{\parbox{2.0cm}{\centering \hspace{2.5mm}Social\newline Connection}} 
& Q16: I feel like \textbf{watching together} with other viewers. & 6.58 & 0.90 & 4.50 & 1.88 & $Z=-2.83,\ p<.01$ \\
& Q17: \textbf{Closeness with other viewers} (using IOS Scale \cite{aron1992inclusion}). & 5.50 & 1.17 & 3.83 & 1.53 & $Z=-2.55,\ p<.05$ \\
& Q18: I feel \textbf{joyful} when viewing videos using the system. & 6.83 & 0.39 & 5.17 & 2.21 & $Z=-2.23,\ p<.05$ \\
\hlineblack
\end{tabular}
}
}
\end{table*}

%% file: Tables/trials.tex
\begin{table}[htbp]
\centering
\caption{Details of the nine uncompleted trials with the baseline. (All 36 trials were completed with DanmuA11y.)}
\renewcommand{\arraystretch}{1.3}
\small
\begin{tabular}{c|p{1cm}|p{1cm}|p{1.5cm}|p{1cm}|p{1cm}|p{1cm}}
\toprule
\textbf{Uncompleted Videos} & \textbf{V3} & \textbf{V4} & \textbf{V5} & \textbf{V6} & \textbf{V7} & \textbf{V8} \\
\midrule
\textbf{Participants} & P15 & P9, P17 & P10, P15, P18 & P12, P17 & - & P20 \\
\bottomrule
\end{tabular}
\label{tab:trials}
\end{table}